\documentclass[sigconf]{acmart}

\usepackage{booktabs}
\usepackage{graphicx, epstopdf}
\usepackage{subfigure}
\usepackage{url}
\usepackage{multirow}
\usepackage{algorithm}
\usepackage{algorithmic}
\usepackage{balance}
\usepackage{arydshln}

\renewcommand*{\textrightarrow}{$\rightarrow$}

\begin{document}

\title{Learning Graph Meta Embeddings for Cold-Start Ads in Click-Through Rate Prediction}
\author{Wentao Ouyang, Xiuwu Zhang, Shukui Ren, Li Li, Kun Zhang, Jinmei Luo, Zhaojie Liu, Yanlong Du}
\affiliation{%
  \institution{Alibaba Group}
}
\email{{maiwei.oywt, xiuwu.zxw, shukui.rsk, ll98745, jerry.zk, cathy.jm, zhaojie.lzj, yanlong.dyl}@alibaba-inc.com}

\begin{abstract}
Click-through rate (CTR) prediction is one of the most central tasks in online advertising systems. Recent deep learning-based models that exploit feature embedding and high-order data nonlinearity have shown dramatic successes in CTR prediction. However, these models work poorly on cold-start ads with new IDs, whose embeddings are not well learned yet.
In this paper, we propose Graph Meta Embedding (GME) models that can rapidly learn how to generate desirable initial embeddings for new ad IDs based on graph neural networks and meta learning. Previous works address this problem from the new ad itself, but ignore possibly useful information contained in existing old ads. In contrast, GMEs simultaneously consider two information sources: the new ad and existing old ads. For the new ad, GMEs exploit its associated attributes. For existing old ads, GMEs first build a graph to connect them with new ads, and then adaptively distill useful information. We propose three specific GMEs from different perspectives to explore what kind of information to use and how to distill information. In particular, GME-P uses \underline{P}re-trained neighbor ID embeddings, GME-G uses \underline{G}enerated neighbor ID embeddings and GME-A uses neighbor \underline{A}ttributes. Experimental results on three real-world datasets show that GMEs can significantly improve the prediction performance in both cold-start (i.e., no training data is available) and warm-up (i.e., a small number of training samples are collected) scenarios over five major deep learning-based CTR prediction models. GMEs can be applied to conversion rate (CVR) prediction as well.
\end{abstract}



\copyrightyear{2021}
\acmYear{2021}
\setcopyright{acmcopyright}\acmConference[SIGIR '21]{Proceedings of the 44th International ACM SIGIR Conference on Research and Development in Information Retrieval}{July 11--15, 2021}{Virtual Event, Canada}
\acmBooktitle{Proceedings of the 44th International ACM SIGIR Conference on Research and Development in Information Retrieval (SIGIR '21), July 11--15, 2021, Virtual Event, Canada}
\acmPrice{15.00}
\acmDOI{10.1145/3404835.3462879}
\acmISBN{978-1-4503-8037-9/21/07}

\settopmatter{printacmref=true}
\fancyhead{}

\maketitle

\section{Introduction}
Click-through rate (CTR) prediction plays an important role in online advertising systems. It aims to predict the probability that a user will click on a specific ad. The predicted CTR impacts both the ad ranking strategy and the ad charging model \cite{zhou2018deep,ouyang2019deep}. For example, the ad ranking strategy generally depends on CTR $\times$ bid, where bid is the benefit the system receives if an ad is clicked. Moreover, according to the cost-per-click (CPC) or the optimized cost-per-click (oCPC) charging model, advertisers are only charged once their ads are clicked by users.
Therefore, in order to maintain a desirable user experience and to maximize the revenue, it is crucial to estimate the CTR accurately.

CTR prediction has attracted lots of attention from both academia and industry \cite{he2014practical,cheng2016wide,zhang2016deep,wang2017deep,zhou2018deep,ouyang2019click,song2019autoint,qin2020user}. In recent years, deep learning-based models such as Deep Neural Network (DNN) \cite{cheng2016wide}, Product-based Neural Network (PNN) \cite{qu2016product}, Wide\&Deep \cite{cheng2016wide}, DeepFM \cite{guo2017deepfm}, xDeepFM \cite{lian2018xdeepfm} and AutoInt \cite{song2019autoint} are proposed to automatically learn latent feature representations and complicated feature interactions in different manners. These models generally follow an Embedding and Multi-layer perceptron (MLP) paradigm, where an embedding layer transforms each raw input feature into a dense real-valued vector representation in order to capture richer semantics and to overcome the limitations of one-hot encoding \cite{mikolov2013distributed}.

Despite the remarkable success of these models, it is extremely data demanding to well learn the embedding vectors.
It has been widely known that a well-learned embedding for an ad ID can largely improve the CTR prediction accuracy \cite{cheng2016wide,guo2017deepfm,qu2016product,juan2016field,zhou2018deep,ouyang2019deep}. When a new ad is added to the candidate pool, its ID is never seen in the training data and therefore no embedding vector is available. A randomly generated ID embedding is unlikely to lead to good prediction performance.
Moreover, for ads with a small number of training samples, it is hard to train their embeddings as good as those with abundant training data. These difficulties are known as the cold-start problem in CTR prediction.

In the domain of cold-start recommendation, some methods propose to use side information, e.g., user attributes \cite{roy2016latent,seroussi2011personalised,zhang2014addressing} and/or item attributes \cite{saveski2014item,schein2002methods,vartak2017meta}.
However, in the CTR prediction task, side information is already used. The aforementioned CTR prediction models are all feature-rich models, which already take user attributes and ad attributes as input.

Another possible way to tackle this problem is to actively collect more training data in a short time. For example, \cite{li2010contextual,nguyen2014cold,shah2017practical,tang2015personalized} use contextual-bandit approaches and \cite{golbandi2011adaptive,harpale2008personalized,park2006naive,zhou2011functional} design interviews to collect specific information with active learning. However, these approaches still cannot lead to satisfactory prediction performance before sufficient training data are collected.

We tackle the cold-start problem for new ads from a different perspective, which is to generate desirable initial embeddings for new ad IDs in a meta learning framework, even when the new ads have no training data at all.
Along this line, Pan et al. propose the Meta-Embedding model \cite{pan2019warm} by exploiting the associated attributes of the new ad.
However, this model only considers the new ad itself, but ignores possibly useful information contained in existing old ads that may help boost the prediction performance. Another meta learning-based model MeLU \cite{lee2019melu} is proposed to estimate a new user's preferences with a few consumed items. It locally updates a user's decision-making process based on the user's item-consumption pattern. This model does not apply to our problem and it also considers the target user alone.

In this paper, we propose Graph Meta Embedding (GME) models to learn how to generate desirable initial embeddings for new ad IDs based on graph neural networks and meta learning.
GMEs contain two major components: 1) embedding generator (EG) and 2) graph attention network (GAT) \cite{velivckovic2018graph}, where the aim of EG is to generate an ID embedding and the aim of GAT is to adaptively distill information.
The main idea of GMEs is to simultaneously consider two information sources: 1) the new ad itself and 2) existing old ads. For the new ad, GMEs exploit its associated attributes. For existing old ads, GMEs first build a graph to connect them with new ads, and then utilize the GAT to adaptively distill useful information. This process is non-trivial, and we propose three specific GMEs from different perspectives. In particular, GME-P uses \underline{P}re-trained neighbor ID embeddings, GME-G uses \underline{G}enerated neighbor ID embeddings and GME-A uses neighbor \underline{A}ttributes. In other words, although the three GME models all exploit the GAT, they differ in what kind of information to use and how to distill information.

In order to train GMEs, we use a gradient-based meta learning approach \cite{pan2019warm}, which generalizes Model-Agnostic Meta-Learning (MAML) \cite{finn2017model}. We view the learning of ID embedding of each ad as a task. We use meta learning because the number of unique ads is much smaller than the number of samples and we need fast adaptation. The loss function considers two aspects: 1) cold-start phase: when a new ad comes in, one should make predictions with a small loss and 2) warm-up phase: after observing a small number of labeled samples, one should speed up the model fitting to reduce the loss for subsequent prediction. As a result, GMEs can improve the CTR prediction performance on new ads in both the cold-start phase (i.e., no training data is available) and the warm-up phase (i.e., a small number of training samples are collected).

The main contributions of this work are summarized as follows:
\begin{itemize}
\item We address the cold-start CTR prediction problem for new ads from a different perspective. The main idea is to build an ad graph and learn to generate desirable initial embeddings for new ad IDs by taking into account of both the new ad itself and other related old ads over the graph.
\item We propose three specific Graph Meta Embedding (GME) models to generate initial embeddings for new ad IDs from different perspectives. In particular, GME-P uses \underline{P}re-trained neighbor ID embeddings, GME-G uses \underline{G}enerated neighbor ID embeddings and GME-A uses neighbor \underline{A}ttributes. We make the implementation code publicly available\footnote{https://github.com/oywtece/gme}.
\item We conduct experiments on three large-scale real-world datasets. Experimental results show that GMEs can significantly improve the prediction performance in both cold-start and warm-up scenarios over five major deep learning-based CTR prediction models.
\end{itemize}

\begin{table}[!t]
\caption{Each row is an instance for CTR prediction. The first column is the label (1 - clicked, 0 - unclicked). Each of the other columns is a field. Instantiation of a field is a feature.}
\vskip -8pt
\label{tab_ft}
\centering
\begin{tabular}{|c|c|c|c|c|}
\hline
\textbf{Label} & \textbf{User ID} & \textbf{User Age} & \textbf{Ad Title} \\
\hline
1 & 2135147 & 24 & Beijing flower delivery \\
\hline
0 & 3467291 & 31 & Nike shoes, sporting shoes \\
\hline
0 & 1739086 & 45 & Female clothing and jeans \\
\hline
\end{tabular}
\end{table}

\section{Background}

\subsection{Problem Formulation}
The task of \textbf{CTR prediction} in online advertising is to build a prediction model to estimate the probability of a user clicking on a specific ad.
Each instance can be described by multiple \emph{fields} such as user information (``User ID'', ``City'', ``Age'', etc.) and ad information (``Ad ID'', ``Category'', ``Title'', etc.). The instantiation of a field is a \emph{feature}. For example, the ``User ID'' field may contain features such as ``2135147'' and ``3467291''. Table \ref{tab_ft} shows some examples.

During model training, we learn parameters corresponding to training features. After that, we make predictions on test data by using these learned parameters. However, new ads have features that are not seen in the training data, e.g., the ad ID. This causes the \textbf{cold-start CTR prediction} problem of new ads, where we need to make predictions in the absence of certain model parameters.
We will make this problem more concrete in the context of deep CTR prediction models introduced below.

\subsection{Typical Deep CTR Prediction Models} \label{sec_dnn}

\begin{figure}[!t]
\centering
\includegraphics[width=0.36\textwidth, trim = 0 0 0 0, clip]{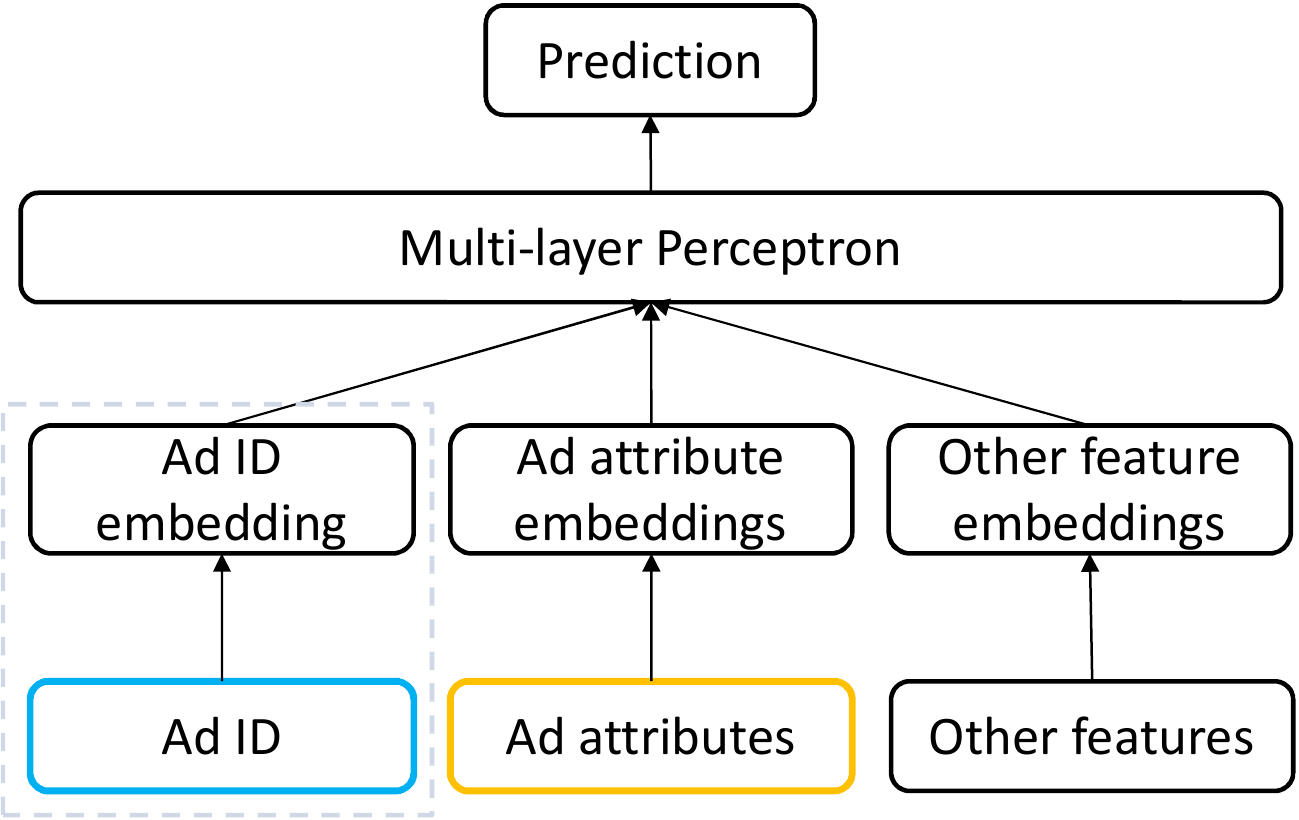}
\vskip -6pt
\caption{Illustration of typical deep CTR prediction models.}
\vskip -8pt
\label{ctr_model}
\end{figure}

Typical deep learning-based CTR prediction models such as Deep Neural Network (DNN) \cite{cheng2016wide}, Product-based Neural Network (PNN) \cite{qu2016product}, Wide\&Deep \cite{cheng2016wide}, DeepFM \cite{guo2017deepfm}, xDeepFM \cite{lian2018xdeepfm} and AutoInt \cite{song2019autoint} all follow an Embedding and MLP paradigm (Figure \ref{ctr_model}). We present the modules of DNN below as an example.

\textbf{Input}: The input to the model is feature indices $\{i\}$.

\textbf{Embedding layer}: $i \rightarrow \mathbf{e}_i$. This module encodes the input into dense vector representations (i.e., embeddings) $\mathbf{e}_i$ through an embedding matrix $\mathbf{E}$ (to be learned). The $i$th column of the embedding matrix $\mathbf{E}$ holds the embedding vector for the $i$th feature.
The embedding vector $\mathbf{e}_i$ for feature index $i$ is given by $\mathbf{e}_i = \mathbf{E}[:,i]$.

\textbf{Concatenation layer}: $\{\mathbf{e}_i\} \rightarrow \mathbf{s}$. This module concatenates the embeddings of all the input features as a long embedding vector $\mathbf{s} = [\mathbf{e}_1 \| \mathbf{e}_2 \| \mathbf{e}_3 \| \cdots]$, where $\|$ is the concatenation operator.

\textbf{Hidden layers}: $\mathbf{s} \rightarrow \mathbf{s}'$. This module transforms the long embedding vector $\mathbf{s}$ into a high-level representation vector $\mathbf{s}'$ through several fully-connected (FC) layers to exploit data nonlinearity and high-order feature interactions. In particular, $\mathbf{s}' = f_L(\cdots f_2((f_1(\mathbf{s}))))$, where $L$ is the number of FC layers and $f_j$ is the $j$th FC layer.

\textbf{Prediction layer}: $\mathbf{s}' \rightarrow \hat{y}$. This module predicts the click-through probability $\hat{y} \in [0, 1]$ of the instance based on the high-level representation vector $\mathbf{s}'$ through a sigmoid function.

\textbf{Model training}: The model parameters are learned through the cross-entropy loss on a training data set $\mathbb{Y}$. The loss function is
\begin{equation} \label{loss}
loss = \frac{1}{|\mathbb{Y}|}\sum_{y \in \mathbb{Y}} [ - y \log \hat{y} - (1 - y) \log (1 - \hat{y})],
\end{equation}
where $y\in\{0,1\}$ is the true label corresponding to $\hat{y}$.

When a new ad comes in, its ID has not been trained yet and the model cannot find its embedding in the embedding matrix. In order to predict the CTR, a commonly used approach is to randomly generated an embedding for the new ad ID. However, this approach usually leads to poor prediction performance.

\section{Model Design}
In the following, we propose Graph Meta Embedding (GME) models to learn how to generate desirable initial embeddings (i.e., not random) for new ad IDs based on graph neural networks and meta learning. These initial embeddings can lead to improved CTR prediction performance in both cold-start and warm-up scenarios.

\begin{table}[!t]
\caption{List of notations.}
\vskip -8pt
\label{tab_nota}
\centering
\begin{tabular}{|c|c|}
\hline
\textbf{Notation} & \textbf{Meaning} \\
\hline
$ID_0$ & ID of the new ad \\
\hline
$\mathbf{x}_0$ & associated attributes of the new ad\\
\hline
$\mathbf{z}_0$ & concatenated embedding vector acc. to $\mathbf{x}_0$ \\
\hline
$\tilde{\mathbf{z}}_0$ & refined embedding vector w.r.t. $\mathbf{z}_0$ \\
\hline
$\mathbf{g}_0$ & generated (preliminary) ID emb. of the new ad \\
\hline
$\mathbf{r}_0$ & initial ID emb. of the new ad in CTR prediction\\
\hline
\hline
$ID_i$ & ID of the $i$th ngb. ($i=1, \cdots, N$) \\
\hline
$\mathbf{x}_i$ & associated attributes of the $i$th ngb. \\
\hline
$\mathbf{z}_i$ & concatenated embedding vector acc. to $\mathbf{x}_i$ \\
\hline
$\mathbf{p}_i$ & pre-trained ID embedding of the $i$th ngb. \\
\hline
$\mathbf{g}_i$ & generated ID embedding of the $i$th ngb. \\
\hline
\end{tabular}
\vskip -8pt
\end{table}

\subsection{Overview}
For ease of presentation, we list the notations used in Table \ref{tab_nota}.

\begin{figure}[!t]
\centering
\includegraphics[width=0.42\textwidth, trim = 0 0 0 0, clip]{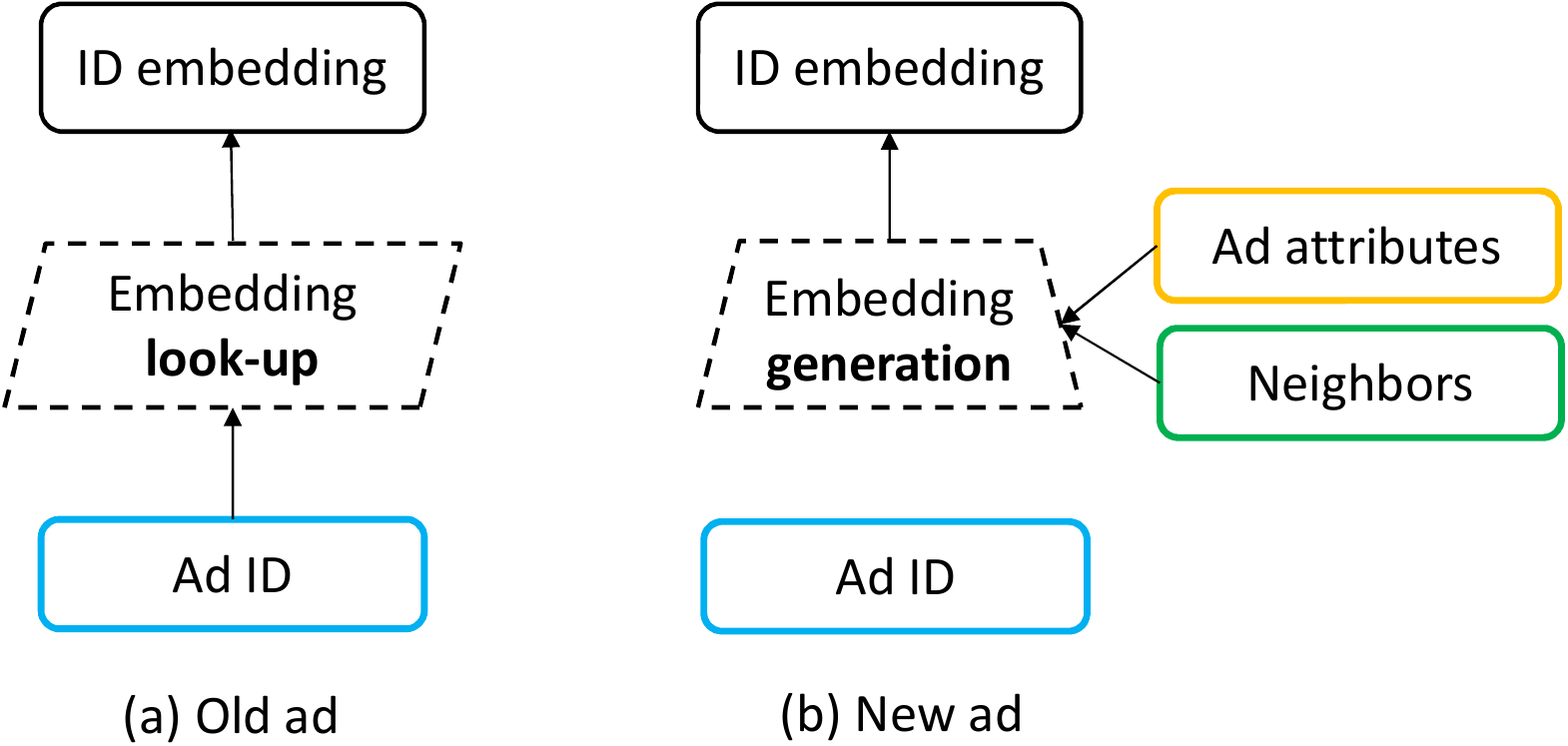}
\vskip -6pt
\caption{Illustration of ID embedding of old and new ads.}
\vskip -4pt
\label{old_new_ad_id}
\end{figure}

The proposed GME models are only activated for new ads. We illustrate the difference of ID embeddings for old ads and new ads in Figure \ref{old_new_ad_id}. When an ad ID is given, we first lookup the trained embedding matrix. If the ID's embedding can be found, then it is an old ad and we use the found embedding [Figure \ref{old_new_ad_id}(a)]. Otherwise, it is a new ad and we activate a GME model to generate an initial embedding for the ID by using the attributes of the new ad and information from its graph neighbors [Figure \ref{old_new_ad_id}(b)].

GMEs contain two major components: 1) embedding generator (EG) and 2) graph attention network (GAT) \cite{velivckovic2018graph}, where the aim of EG is to generate an ID embedding and the aim of GAT is to adaptively distill information. GME models differ in what kind of information to use and how to distill information.

For convenience, we use the same notations for model parameters ($\mathbf{W}$, $\mathbf{V}$ and $\mathbf{a}$) in the following. Parameters with the same notation in different models have the same functionality, but possibly different dimensions (which are clear in each specific context).

\subsection{Graph Creation}

\begin{figure}[!t]
\centering
\includegraphics[width=0.26\textwidth, trim = 0 0 0 0, clip]{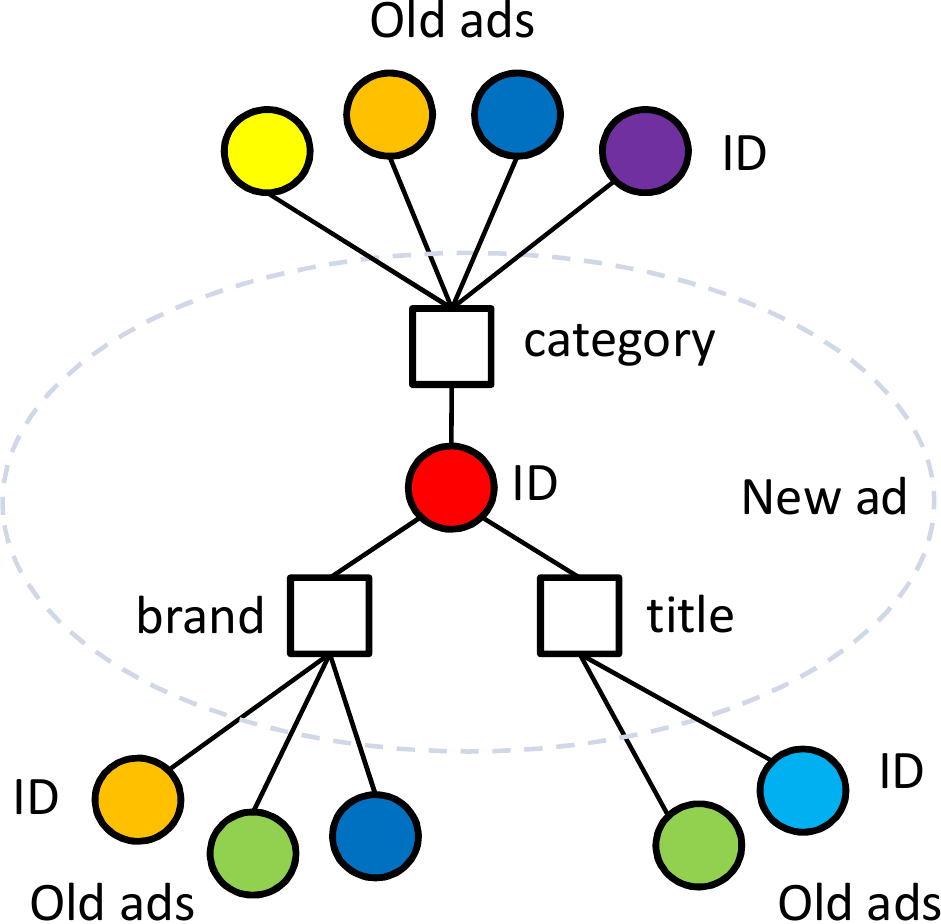}
\vskip -6pt
\caption{Illustration of the ad graph. Connections between ads are established based on their shared features. The dashed circle illustrates an ad with the ID and three attribute features (e.g., category, brand and title) shown.}
\vskip -8pt
\label{graph}
\end{figure}

As the GME models utilize both the new ad and related old ads, the first step is to build a graph to connect them.
However, unlike social networks where exist follower-followee or friendship relationships, there is no natural graph on ads. One possible way is to exploit the co-click relationships, but this approach is clearly not suitable for new ads. In this paper, we build connections between ads based on their features (illustrated in Figure \ref{graph}).

Typically, one can use an adjacency matrix $\mathbf{A}$ \cite{chanpuriya2020infinitewalk}, where the $i$th row represents a new ad $i$, the $j$th column represents an existing old ad $j$ and $[\mathbf{A}]_{ij}$ is the adjacency score between $i$ and $j$. This approach is highly time-consuming because it needs to repeatedly scan the whole set of existing old ads for each new ad.

Instead, we use the following approach for fast graph creation. Given a new ad, we obtain its ID and associated attributes (e.g., category, brand and title). For each attribute, we can retrieve old ads which have the same attribute. The union of these old ads then form the graph neighbors of the new ad.

\begin{figure}[!t]
\centering
\includegraphics[width=0.42\textwidth, trim = 0 0 0 0, clip]{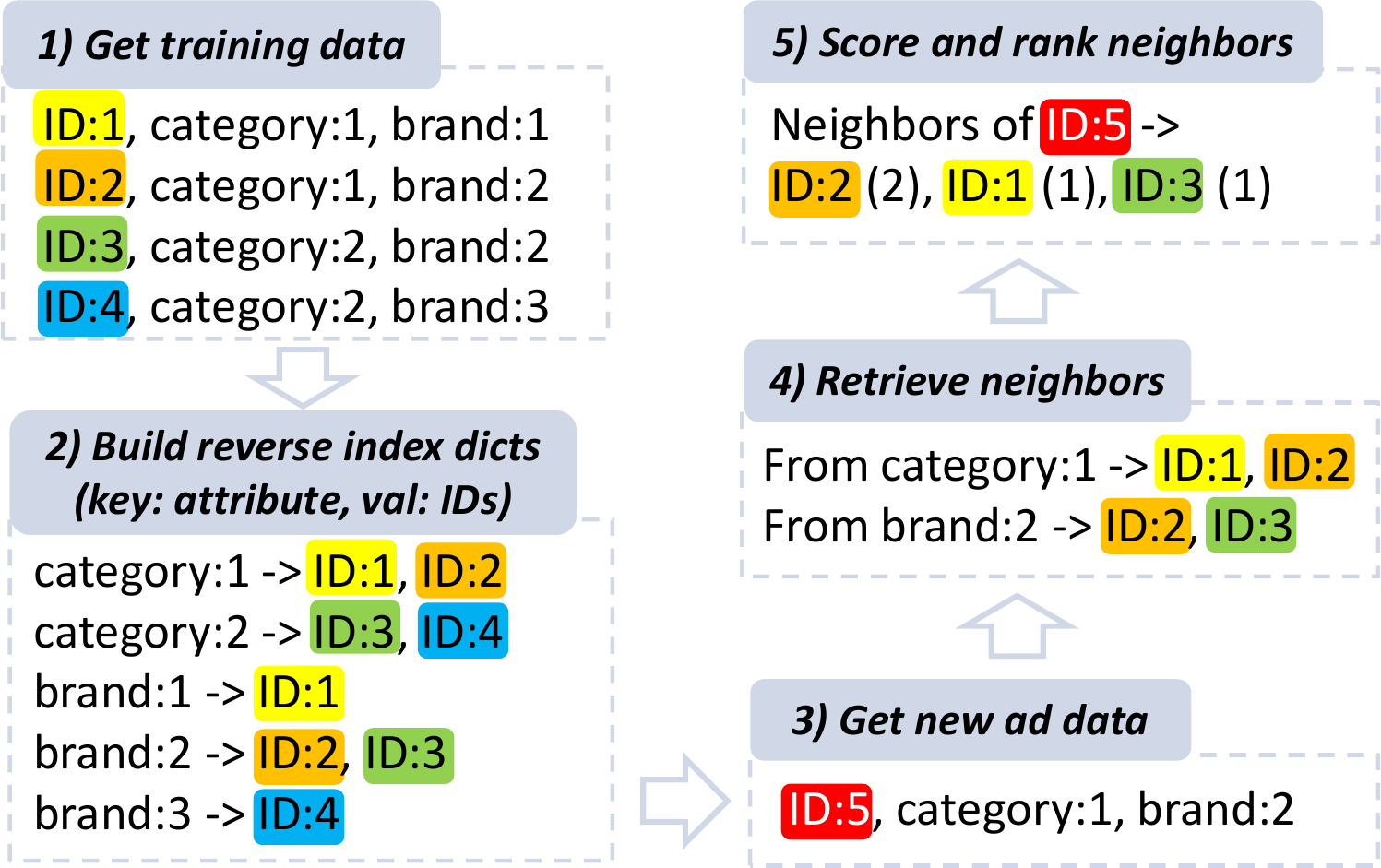}
\vskip -6pt
\caption{Illustration of fast graph creation.}
\vskip -8pt
\label{gen_ngb}
\end{figure}

In particular, we implement this idea as follows (summarized in Figure \ref{gen_ngb}, where Steps 1-2 are performed only once for old ads).
\begin{itemize}
\item Build a reverse index dictionary, where the key is the attribute and the value is the set of ad IDs which have this attribute. For example, ``category:1 $\rightarrow$ ID:1, ID:2''; ``brand:2 $\rightarrow$ ID:2, ID:3''.
\item Given a new ad, retrieve its neighbors based on each attribute. For example, the new ad is ``ID:5, category:1, brand:2'', we then retrieve its neighbors based on the two attributes. The retrieved neighbors are ID:1, ID:2 and ID:3.
\item Calculate the similarity score w.r.t. each neighbor and keep the top-$N$ neighbors (break the tie randomly). We define the score as the number of attributes that the neighbor can be retrieved from. For example, ID:2 has a score as 2 because it can be retrieved from 2 attributes. This step can keep the most useful neighbors for subsequent processing.
\end{itemize}

The above approach is much faster because we only need to scan the set of old ads once (to build the reserve index dictionary) instead of multiple times.
Without loss of generality, we denote the new ad as $ID_0$ and the set of its graph neighbors as $\mathcal{N} = \{ID_i\}_{i=1}^N$.

\subsection{GME-P: Using \underline{P}re-trained Neighbor ID Embeddings} \label{sec_gme_pre}
The first GME model we present is GME-P, which exploits the attributes of the new ad and the pre-trained ID embeddings of neighboring old ads. We illustrate its structure in Figure \ref{model_gme}(a).

The idea is that: As we have obtained the pre-trained ID embeddings $\{\mathbf{p}_i\}$ of neighboring old ads by the main CTR prediction model (e.g. DNN in \S\ref{sec_dnn}), we would like to exploit useful information in these embeddings. However, we only have attribute embeddings rather than ID embedding of the new ad. Therefore, we first generate a preliminary ID embedding $\mathbf{g}_0$ for the new ad by using its associated attributes. As both $\{\mathbf{p}_i\}$ and $\mathbf{g}_0$ are ID embeddings, we can then leverage useful information contained in $\{\mathbf{p}_i\}$ to improve $\mathbf{g}_0$ and obtained a refined ID embedding $\mathbf{r}_0$ for the new ad. Formally, GME-P contains the following two steps.

\subsubsection{\textbf{ID Embedding Generation}}
We generate a preliminary ID embedding for a cold-start ad by using its associated attribute features (such as category, brand and title) instead of randomly. Formally, let's denote the features of an instance as $[ID_0, \mathbf{x}_0, \mathbf{o}_0]$, where $ID_0$ is the identity of the new ad, $\mathbf{x}_0$ is the ad attribute features, and $\mathbf{o}_0$ is other features which do not necessarily relate to the ad such as user features and context features.

Although $ID_0$ of the new ad is not seen in the training data, the associated ad attributes $\mathbf{x}_0$ are usually observed.
We then lookup the embeddings corresponding to $\mathbf{x}_0$ and obtain a long concatenated embedding vector $\mathbf{z}_0$.
Based on $\mathbf{z}_0$, we generate a preliminary embedding $\mathbf{g}_0$ for $ID_0$ through an embedding generator (EG) which implements $\mathbf{g}_0 = f(\mathbf{z}_0)$. We use a simple instantiation of the EG as
\begin{equation} \label{eq_meta_new}
\mathbf{g}_0 = \gamma \tanh(\mathbf{W z}_0),
\end{equation}
where $\mathbf{W}$ is the parameter (to be learned) of a fully connected layer, $\tanh$ is the activation function and $\gamma \in (0,1]$ is a scaling hyperparameter. We use $\gamma$ to restrict the range of $\mathbf{g}_0$ in $[-\gamma, \gamma]$.

\subsubsection{\textbf{ID Embedding Refinement}}
We then generate a refined ID embedding $\mathbf{r}_0$ for the new ad based on its preliminary ID embedding $\mathbf{g}_0$ and pre-trained ID embeddings $\{\mathbf{p}_i\}$ of its neighbors.

A simple way is to take the average of these ID embeddings and obtain $\mathbf{r}_0 = \textnormal{average}(\mathbf{g}_0, \mathbf{p}_1, \mathbf{p}_2, \cdots, \mathbf{p}_N)$. But this is clearly not a wise choice because some old ads may not be quite informative.

Alternatively, we resort to the Graph Attention Network (GAT) \cite{velivckovic2018graph}, which is proposed to operate on graph-structured data and to learn high-level data representations. It allows for assigning different importances to different graph nodes within a neighborhood through the attention mechanism \cite{bahdanau2014neural} while dealing with different sized neighborhoods.

\begin{figure*}[!t]
\centering
\includegraphics[width=0.98\textwidth, trim = 0 0 0 0, clip]{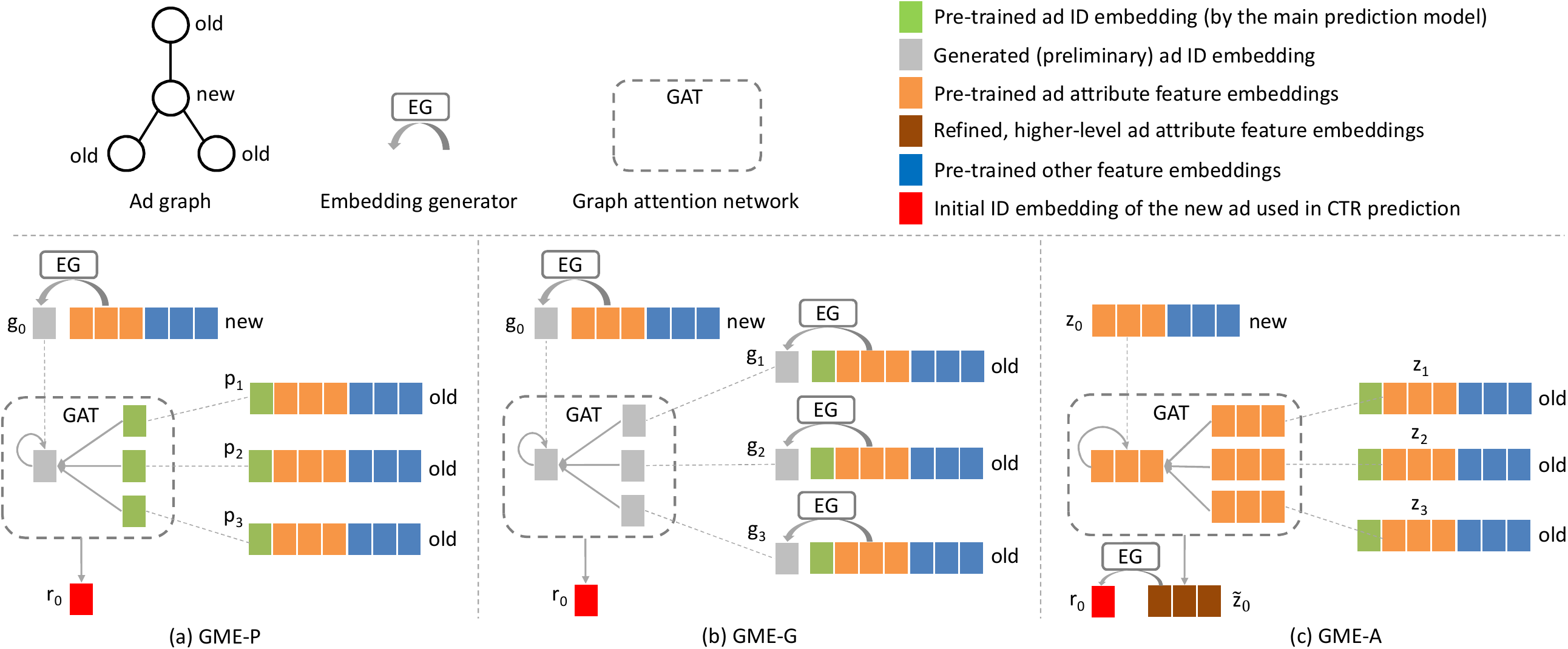}
\vskip -6pt
\caption{Graph Meta Embedding (GME) models (best viewed in color). Please refer to \S\ref{sec_gme_pre} to \S\ref{sec_gme_attr} for more details.}
\vskip -5pt
\label{model_gme}
\end{figure*}

We first compute the attention coefficient between $\mathbf{g}_0$ and $\mathbf{p}_i$ as
\[
c_{0i} = \mathcal{F}(\mathbf{V g}_0, \mathbf{V p}_i),
\]
where $\mathcal{F}$ is a function to implement the attention mechanism and $\mathbf{V}$ is a shared weight parameter which transforms the input into higher-level features and obtains sufficient expressive power.
We also compute the attention coefficient for the new ad itself as
\[
c_{00} = \mathcal{F}(\mathbf{V g}_0, \mathbf{V g}_0).
\]

To make coefficients easily comparable across different nodes, we normalize them using the softmax function.
We implement the attention mechanism $\mathcal{F}$ using a single-layer feedforward neural network, parameterized by a weight vector $\mathbf{a}$, and applying the LeakyReLU nonlinearity (with negative input slope 0.2) \cite{velivckovic2018graph}. The normalized coefficients $\alpha_{0i}$ can then be expressed as
\[
\alpha_{0i} = \frac{\exp(c_{0i})}{\sum_{j=0}^N \exp(c_{0j})} = \frac{\exp\left(LeakyReLU(\mathbf{a}^T [\mathbf{V g}_0 \| \mathbf{V p}_i]) \right)}{\sum_{j=0}^N \exp\left(LeakyReLU (\mathbf{a}^T [\mathbf{V g}_0 \| \mathbf{V p}_j]) \right)},
\]
where we define $\mathbf{p}_0 \triangleq \mathbf{g}_0$ for notational simplicity. LeakyReLU allows to encode both positive and small negative signals \cite{maas2013rectifier}.

Note that the index $j$ ranges from 0 to $N$. That is, the summation includes the new ad itself (index 0) and its neighbors (index 1 to $N$).

We then compute a weighted sum of the preliminary ID embedding $\mathbf{g}_0$ of the new ad (with importance $\alpha_{00}$) and the pre-trained ID embeddings $\{\mathbf{p}_i\}$ of neighbors (with importance $\alpha_{0i}$), to serve as the refined ID embedding $\mathbf{r}_0$ for the new ad as
\[
\mathbf{r}_0 = ELU \left(\sum_{i=0}^N \alpha_{0i} \mathbf{V p}_i \right),
\]
where ELU is the exponential linear unit activation function \cite{velivckovic2018graph}, it also allows to encode both positive and small negative
signals.

\subsubsection{\textbf{Analysis}}
GME-P seems reasonable. However, the pre-training process (e.g., based on the DNN model) does not impose any constraint between attributes and the ID embedding as in Eq. (\ref{eq_meta_new}) and all the embeddings are randomly initialized. It is possible that given the same attributes, the corresponding $\mathbf{p}_0$ and $\mathbf{g}_i$ are quite different (because they correspond to two different IDs). It makes the attention computation between $\mathbf{p}_0$ and $\mathbf{g}_i$ meaningless.

\subsection{GME-G: Using \underline{G}enerated Neighbor ID Embeddings} \label{sec_gme_gen}
To overcome the limitation of GME-P, we propose GME-G in this section. Instead of using pre-trained ID embeddings of old ads, GME-G reuses the EG for the new ad, and generates ID embeddings $\{\mathbf{g}_i\}$ for old ads using their corresponding attributes as well. We illustrate its structure in Figure \ref{model_gme}(b).

\subsubsection{\textbf{ID Embedding Generation}}
We use the same EG to generate the preliminary ID embedding $\mathbf{g}_0$ for the new ad and the ID embeddings $\{\mathbf{g}_i\}$ for existing old ads as
\[
\mathbf{g}_0 = \gamma \tanh(\mathbf{W z}_0), \ \mathbf{g}_i = \gamma \tanh(\mathbf{W z}_i).
\]

By doing so, we can guarantee that when ad attributes are the same (i.e., $\mathbf{z}_i = \mathbf{z}_0$), we have $\mathbf{g}_i = \mathbf{g}_0$.
Subsequently, the attention computation between $\mathbf{g}_i$ and $\mathbf{g}_0$ makes more sense.

\subsubsection{\textbf{ID Embedding Refinement}}
The attention coefficients between the new ad and the $i$th neighboring old ad is then given by
\[
\alpha_{0i} = \frac{\exp \left(LeakyReLU(\mathbf{a}^T [\mathbf{V g}_0 \| \mathbf{V g}_i]) \right)}{\sum_{j=0}^N \exp \left(LeakyReLU(\mathbf{a}^T [\mathbf{V g}_0 \| \mathbf{V g}_j]) \right)}.
\]

Finally, the refined ID embedding $\mathbf{r}_0$ for the new ad is given by a linear combination of the generated ID embedding of the new ad and those of the neighboring old ads as
\[
\mathbf{r}_0 = ELU \left(\sum_{i=0}^N \alpha_{0i} \mathbf{V g}_i \right).
\]

\subsubsection{\textbf{Analysis}}
GME-G does overcome the limitation of GME-P and it makes the attention coefficients between the new ad and old ads meaningful. However, GME-G repeatedly performs ID embedding generation for old ads. As the generated ID embedding could contain certain noise, the repetition can spread the noise.

\subsection{GME-A: Using Neighbor \underline{A}ttributes} \label{sec_gme_attr}
Given the limitation of GME-G, we further propose GME-A in this section, whose structure is shown in Figure \ref{model_gme}(c). GME-A reverses the order of the ``generation'' step and the ``refinement'' step. Moreover, GME-A refines the attribute representation rather than the preliminary ID embedding.

\subsubsection{\textbf{Attribute Embedding Refinement}}
GME-A first obtains a refined attribute representation of the new ad, which aggregates useful information from the new ad itself and its neighboring old ads on the attribute level.
Formally, the attention coefficients between the new ad and the $i$th neighboring old ad is computed based on attribute embedding vectors $\mathbf{z}_0$ and $\mathbf{z}_i$ as
\[
\alpha_{0i} = \frac{\exp \left(LeakyReLU(\mathbf{a}^T [\mathbf{V z}_0 \| \mathbf{V z}_i])\right)}{\sum_{j=0}^N \exp \left(LeakyReLU(\mathbf{a}^T [\mathbf{V z}_0 \| \mathbf{V z}_j]) \right)}.
\]

We then obtain a refined, high-level attribute embedding vector $\tilde{\mathbf{z}}_0$ for the new ad by performing a linear combination of the original embedding vectors as
\[
\tilde{\mathbf{z}}_0 = ELU \left(\sum_{i=0}^N \alpha_{0i} \mathbf{V z}_i \right).
\]

\subsubsection{\textbf{ID Embedding Generation}}
Given this refined attribute representation, we then generate the initial ID embedding of the new ad as
\[
\mathbf{r}_0 = \gamma \tanh(\mathbf{W} \tilde{\mathbf{z}}_0).
\]

\subsubsection{\textbf{Analysis}}
GME-A directly compares the attributes of the new ad and neighboring old ads, thus avoiding the ``incomparable'' problem between the generated ID embedding and the pre-trained ID embeddings in GME-P. GME-A only uses the EG once, thus also avoiding the ``repetition'' issue in GME-G.

\subsection{Model Learning}
We first train a main model (e.g., DNN) for CTR prediction using old ads. We then obtain the model parameters $\Theta$, including the embedding vectors of features and other weight parameters. As $\Theta$ is usually trained with a large amount of data, we are confident about its effectiveness. Therefore, when training the GME models, we freeze $\Theta$ and only learn the parameters $\Psi \triangleq \{\mathbf{W}, \mathbf{V}, \mathbf{a}\}$ that are specific to these models.

As can be seen, the number of unique ad IDs matters in the training of parameters $\Psi$. As the number of unique ad IDs is much smaller than the number of samples, we resort to meta learning for fast adaptation. We view the learning of ID embedding of each ad as a task and use a gradient-based meta learning approach \cite{pan2019warm}, which generalizes Model-Agnostic Meta-Learning (MAML) \cite{finn2017model}.

The loss consider two aspects: 1) the error of CTR prediction for the new ad should be small and 2) after a small number of labeled examples are collected, a few gradient updates should lead to fast learning. This is achieved by combining the following two losses $l_a$ and $l_b$.

For a given training old ad $ID_0$, we randomly select two disjoint minibatches of labeled data $\mathcal{D}^a$ and $\mathcal{D}^b$, each with $M$ samples. We first make predictions using the initial ID embedding $\mathbf{r}_0$ produced by a GME model on the first minibatch $\mathcal{D}^a$. For the $j$th sample, we obtain its prediction as $\hat{y}_{aj}$.
The average loss over these samples is given by
\[
l_a = \frac{1}{M} \sum_{j=1}^M [-y_{aj} \log \hat{y}_{aj} - (1-y_{aj}) \log(1 - \hat{y}_{aj})],
\]
where $y_{aj}$ is the true label.

Next, by computing the gradient of $l_a$ w.r.t. the initial embedding and taking a step of gradient descent, we get a new adapted embedding
\[
\mathbf{r}'_0 = \mathbf{r}_0 - \eta \frac{\partial l_a}{\partial \mathbf{r}_0},
\]
where $\eta > 0$ is the step size of gradient descent.

We then test this new adapted embedding $\mathbf{r}'_0$ on the second minibatch $\mathcal{D}^b$, and obtain the average loss
\[
l_b = \frac{1}{M} \sum_{j=1}^M [-y_{bj} \log \hat{y}_{bj} - (1-y_{bj}) \log(1 - \hat{y}_{bj})].
\]

The final loss for learning the parameters $\Psi$ is given by
\[
l = \beta l_a + (1-\beta) l_b,
\]
where $\beta \in [0,1]$ is a coefficient to balance the two losses that consider the aforementioned two aspects.

\section{Experiments}

\subsection{Datasets}
We evaluate the performance of the proposed GME models on three real-world datasets, whose statistics are listed in Table \ref{tab_stat}.

\begin{table*}[!th]
\renewcommand{\arraystretch}{1.1}
\caption{Statistics of experimental datasets.}
\vskip -6pt
\label{tab_stat}
\centering
\begin{tabular}{|l|l||p{1cm}|p{2.6cm}||p{1cm}|p{3cm}||p{1cm}|p{1.6cm}|p{1.6cm}|}
\hline
\textbf{Dataset} &\textbf{\# fields} & \textbf{\# old ad IDs} & \textbf{\# samples to train the main prediction model} & \textbf{\# old ad IDs} & \textbf{\# samples to train the cold-start ID embedding model} & \textbf{\# new ad IDs} & \textbf{\# samples for warm up training} & \textbf{\# samples for testing} \\
\hline
ML-1M & 8 & 1,058 & 765,669 & 1,058 & 42,320 & 1,127 & 67,620 & 123,787 \\
\hline
Taobao & 23 & 62,209 & 835,450 & 3,177 & 254,160 & 531,593 & 808,806 & 896,615\\
\hline
News feed & 30 & 5,563 & 3,088,542 & 1,761 & 352,000 & 8,379 & 603,335 & 1,346,504 \\
\hline
\end{tabular}
\end{table*}

(1) \textbf{MovieLens-1M (ML-1M) dataset\footnote{http://www.grouplens.org/datasets/movielens/}.} It is one of the most well-known benchmark dataset. This dataset contains 1 million movie rating instances over thousands of movies and users. Each movie has an ID and can be seen as an ad in our scenario. The associated attribute features include year of release, title and genres. Other features include user ID, gender, age and occupation. We convert the ratings that are at least 4 to label 1 and others to label 0. This is a common practice for evaluation in implicit feedback scenarios such as CTR prediction \cite{he2017neural}.

(2) \textbf{Taobao ad dataset\footnote{https://tianchi.aliyun.com/dataset/dataDetail?dataId=408}.} It is gathered from the traffic logs in Taobao \cite{ma2018entire} and is originally used for the conversion rate (CVR) prediction task.
Each ad has an ID and the associated attribute features include category ID, shop ID, brand ID and intention node ID. Other features include user features and context features such as user ID, gender, age and categorical ID of user profile.

(3) \textbf{News feed ad dataset.} It is gathered from an industrial news feed advertising system and is used for CTR prediction. Each ad has an ID and the associated attribute features include industry ID, source ID, account ID and title. Other features include user features and context features such as user ID, gender, age and OS.

\subsection{Experimental Settings}
\subsubsection{\textbf{Main CTR Prediction Models}}
Because GMEs are model-agnostic (they only generate initial embeddings for new ad IDs), they can be applied upon various existing CTR prediction models that require feature embeddings. We conduct experiments on the following representative CTR prediction models:
\begin{enumerate}
\item \textbf{DNN}. Deep Neural Network in \cite{cheng2016wide}. It contains an embedding layer, several FC layers and an output layer.
\item \textbf{PNN}. Product-based Neural Network in \cite{qu2016product}. It introduces a production layer into DNN.
\item \textbf{Wide\&Deep}. Wide\&Deep model in \cite{cheng2016wide}. It combines logistic regression (LR) and DNN.
\item \textbf{DeepFM}. DeepFM model in \cite{guo2017deepfm}. It combines factorization machine (FM) \cite{rendle2010factorization} and DNN.
\item \textbf{AutoInt}. AutoInt model in \cite{song2019autoint}. It consists of a multi-head self-attentive network with residual connections and DNN.
\end{enumerate}

There are other CTR prediction models that take additional information into consideration. For example, Deep Interest Network (DIN) \cite{zhou2018deep} models user interest based on historical click behavior. Deep Spatio-Temporal Network (DSTN) \cite{ouyang2019deep} jointly exploits contextual ads, clicked ads and unclicked ads for CTR prediction. As most datasets do not contain behavior sequence information or position information, we do not include these models in our experiments.

\subsubsection{\textbf{Cold-Start ID Embedding Models}}
For each main CTR prediction model, we evaluate the following cold-start ID embedding models, which generate initial embeddings for new ad IDs.
\begin{enumerate}
\item \textbf{RndEmb}. It uses a randomly generated embedding for the new ad ID.
\item \textbf{MetaEmb}. MetaEmbedding model in \cite{pan2019warm}. It uses the attributes $\mathbf{x}_0$ of the new ad to generate an initial embedding of the new ad ID. MetaEmb serves as a baseline which only considers the new ad.
\item \textbf{NgbEmb}. It uses pre-trained ID embeddings of neighboring old ads to generate an initial ID embedding of the new ad as $\gamma \tanh(\mathbf{W} \frac{1}{N}\sum_{i=1}^N \mathbf{p}_i)$. NgbEmb serves as a baseline which only considers neighbor information.
\item \textbf{GME-P}. Graph Meta Embedding model which uses \underline{P}re-trained ID embeddings $\{\mathbf{p}_i\}$ of neighboring old ads and the attributes $\mathbf{x}_0$ of the new ad to generate an initial embedding of the new ad ID. It is described in \S\ref{sec_gme_pre}.
\item \textbf{GME-G}. Graph Meta Embedding model which uses \underline{G}enerated ID embeddings $\{\mathbf{g}_i\}$ of the neighboring old ads and the attributes $\mathbf{x}_0$ of the new ad to generate an initial embedding of the new ad ID. It is described in \S\ref{sec_gme_gen}.
\item \textbf{GME-A}. Graph Meta Embedding model which uses the \underline{A}ttributes $\{\mathbf{x}_i\}$ of neighboring old ads and the attributes $\mathbf{x}_0$ of the new ad to generate an initial embedding of the new ad ID. It is described in \S\ref{sec_gme_attr}.
\end{enumerate}

\subsubsection{\textbf{Parameter Settings}}
We set the dimension of the embedding vector for each feature as 10, the balancing parameter as $\beta=0.1$ and the number of graph neighbors for each ad as $N=10$. For an ad ID, if the number of labeled instances is larger than a threshold, we regard it as an old ad. This threshold is set to 300, 40 and 100 for the three datasets respectively. Old ads are used to train the main CTR prediction model. We further sample old ads to train the cold-start ID embedding models, where each old ad has 20, 40 and 100 samples in each minibatch for the three datasets respectively. For the new ads, we hold out a proportion for warm up training (also serve as validation data) and use the remaining for testing. Details are listed in Table \ref{tab_stat}.
All the models are implemented in Tensorflow \cite{abadi2016tensorflow} and optimized by the Adam algorithm \cite{kingma2014adam}. We run each model 3 times and report the average result.

\subsubsection{\textbf{Evaluation Metrics}}
\begin{enumerate}
\item \textbf{AUC}: Area Under the ROC Curve over the test set. It is a widely used metric for CTR prediction. It reflects the probability that a model ranks a randomly chosen positive instance higher than a randomly chosen negative instance. The larger the better. A small improvement in AUC is likely to lead to a significant increase in online CTR \cite{cheng2016wide,guo2017deepfm,zhou2018deep,ouyang2019deep}.
\item \textbf{Loss}: the value of Eq. (\ref{loss}) of the main prediction model over the test set. The smaller the better.
\end{enumerate}

\subsection{Performance Comparison}
\subsubsection{\textbf{Effectiveness in the Cold-Start Phase}}
Table \ref{tab_auc} lists the performance of various ID embedding models based on different CTR prediction models in the cold-start phase. It is observed that MetaEmb performs better than RndEmb, showing that using associated attributes of the new ad can contribute useful information and alleviate the cold-start problem.
NgbEmb sometimes performs better and sometimes performs worse than MetaEmb, showing that simply considering the average of pre-trained neighbor ID embeddings is not quite effective.

GME-P leads to marginal performance improvement or even degraded performance compared with MetaEmb. It is because the pre-trained neighbor ID embeddings and the generated ID embedding from ad attributes are incomparable. As a consequence, GAT in GME-P cannot well extract useful information from neighbors.

In contrast, GME-G performs much better than MetaEmb. Different from GME-P, GME-G uses generated rather than pre-trained neighbor ID embeddings. As the preliminary ID embedding of the new ad is also generated from ad attributes, these embeddings are comparable. GAT can thus distill informative signals from neighbor ID embeddings and improve the new ad's ID embedding. GME-A further outperforms GME-G in most cases. It is because GME-A directly aggregates useful information from the neighbors on the attribute level and avoids the ``repetition'' issue in GME-G.

These results demonstrate that considering neighbor information and appropriately distilling useful information from them could help alleviate the cold-start problem of new ads.

\newcommand{\tabincell}[2]{\begin{tabular}{@{}#1@{}}#2\end{tabular}}

\begin{table}[!t]
\setlength{\tabcolsep}{1.4pt}
\renewcommand{\arraystretch}{1.2}
\caption{Test AUC and Loss. Pred. model: Prediction model. Emb. model: ID embedding generation model. AUC ($\uparrow$) is the larger the better. Loss ($\downarrow$) is the smaller the better. }
\vskip -5pt
\label{tab_auc}
\centering
\begin{tabular}{|p{1.1cm}|p{1.2cm}|c|c||c|c||c|c|}
\hline
\multicolumn{2}{|c|}{} & \multicolumn{2}{|c||}{\textbf{ML-1M}} & \multicolumn{2}{|c||}{\textbf{Taobao}}  & \multicolumn{2}{|c|}{\textbf{News Feed}} \\
\hline
\textbf{Pred. model} & \textbf{Emb. model} & AUC & Loss & AUC & Loss & AUC & Loss \\
\hline
\multirow{5}{*}{\tabincell{l}{DNN}} & RndEmb & 0.7107 & 0.6491 & 0.6289 & .03177 & 0.7350 & .03602 \\
& MetaEmb & 0.7144 & 0.6439 & 0.6291 & .03177 & 0.7362 & .03578\\
& NgbEmb & 0.7131 & 0.6442 & 0.6294 & .03177 & 0.7356 & .03601 \\
& GME-P & 0.7146 & 0.6437 & 0.6295 & .03177 & 0.7358 & .03602 \\
& GME-G & 0.7217 & 0.6389 & 0.6323 & .03172 & 0.7371 & .03562 \\
& GME-A & \textbf{0.7232} & \textbf{0.6368} & \textbf{0.6336} & \textbf{.03168} & \textbf{0.7389} & \textbf{.03553} \\
\hline
\multirow{5}{*}{\tabincell{l}{PNN}} & RndEmb & 0.7162 & 0.6260 & 0.6325 & .03172 & 0.7334 & .03681 \\
& MetaEmb & 0.7164 & 0.6256 & 0.6327 & .03172 & 0.7365 & .03669 \\
& NgbEmb & 0.7163 & 0.6254 & 0.6330 & .03171 & 0.7329 & .03684 \\
& GME-P & 0.7164 & 0.6258 & 0.6330 & .03172 & 0.7352 & .03672 \\
& GME-G & 0.7172 & 0.6261 & 0.6343 & .03166 & 0.7381 & .03623 \\
& GME-A & \textbf{0.7198} & \textbf{0.6233} & \textbf{0.6354} & \textbf{.03161} & \textbf{0.7392} & \textbf{.03617} \\
\hline
\multirow{5}{*}{\tabincell{l}{Wide\&\\Deep}} & RndEmb & 0.7122 & 0.6509 & 0.6305 & .03164 & 0.7368 & .03565 \\
& MetaEmb &0.7149 & 0.6510 & 0.6306 & .03164 & 0.7381 & .03561 \\
& NgbEmb & 0.7125 & 0.6512 & 0.6306 & .03165  & 0.7354 & .03567 \\
& GME-P & 0.7149 & 0.6510 & 0.6306 & .03166 & 0.7375 & .03529 \\
& GME-G & 0.7166 & 0.6487 & 0.6332 & \textbf{.03142} & 0.7404 & .03514 \\
& GME-A & \textbf{0.7179} & \textbf{0.6425} & \textbf{0.6338} & .03143 & \textbf{0.7413} & \textbf{.03503} \\
\hline
\multirow{5}{*}{\tabincell{l}{DeepFM}} &RndEmb & 0.7143 & 0.6462 & 0.6294 & .03174 & 0.7315 &.03584 \\
& MetaEmb & 0.7146  & 0.6484 & 0.6297 & .03171 & 0.7352 & .03538\\
& NgbEmb & 0.7142 & 0.6467 & 0.6299 & .03171 & 0.7321 & .03585 \\
& GME-P & 0.7146 & 0.6478 & 0.6298 & .03175 & 0.7346 & .03541 \\
& GME-G & 0.7195 & 0.6457 & 0.6337 & \textbf{.03157}  & 0.7378 & .03524 \\
& GME-A & \textbf{0.7206} & \textbf{0.6449}  & \textbf{0.6345} & .03160 & \textbf{0.7389} & \textbf{.03517} \\
\hline
\multirow{5}{*}{\tabincell{l}{AutoInt}} &RndEmb & 0.7152 & 0.6322 & 0.6331 & .03193 & 0.7381 & .03685\\
& MetaEmb & 0.7167 & 0.6224 & 0.6336 & .03166 & 0.7401 & .03672 \\
& NgbEmb & 0.7154 & 0.6251 & 0.6335 & .03164 & 0.7377 & .03691 \\
& GME-P & 0.7168 & 0.6262 & 0.6335 & .03167 & 0.7394 & .03676 \\
& GME-G & 0.7204 & 0.6245 & 0.6402 & .03154 & 0.7416 & .03659 \\
& GME-A & \textbf{0.7223} & \textbf{0.6218} & \textbf{0.6411} & \textbf{.03151} & \textbf{0.7432} & \textbf{.03647} \\
\hline
\end{tabular}
\vskip -8pt
\end{table}

\subsubsection{\textbf{Effectiveness in the Warm-up Phase}}
\begin{figure}[!t]
\centering
\subfigure[ML-1M]{\includegraphics[width=0.235\textwidth, trim = 0 0 0 0, clip]{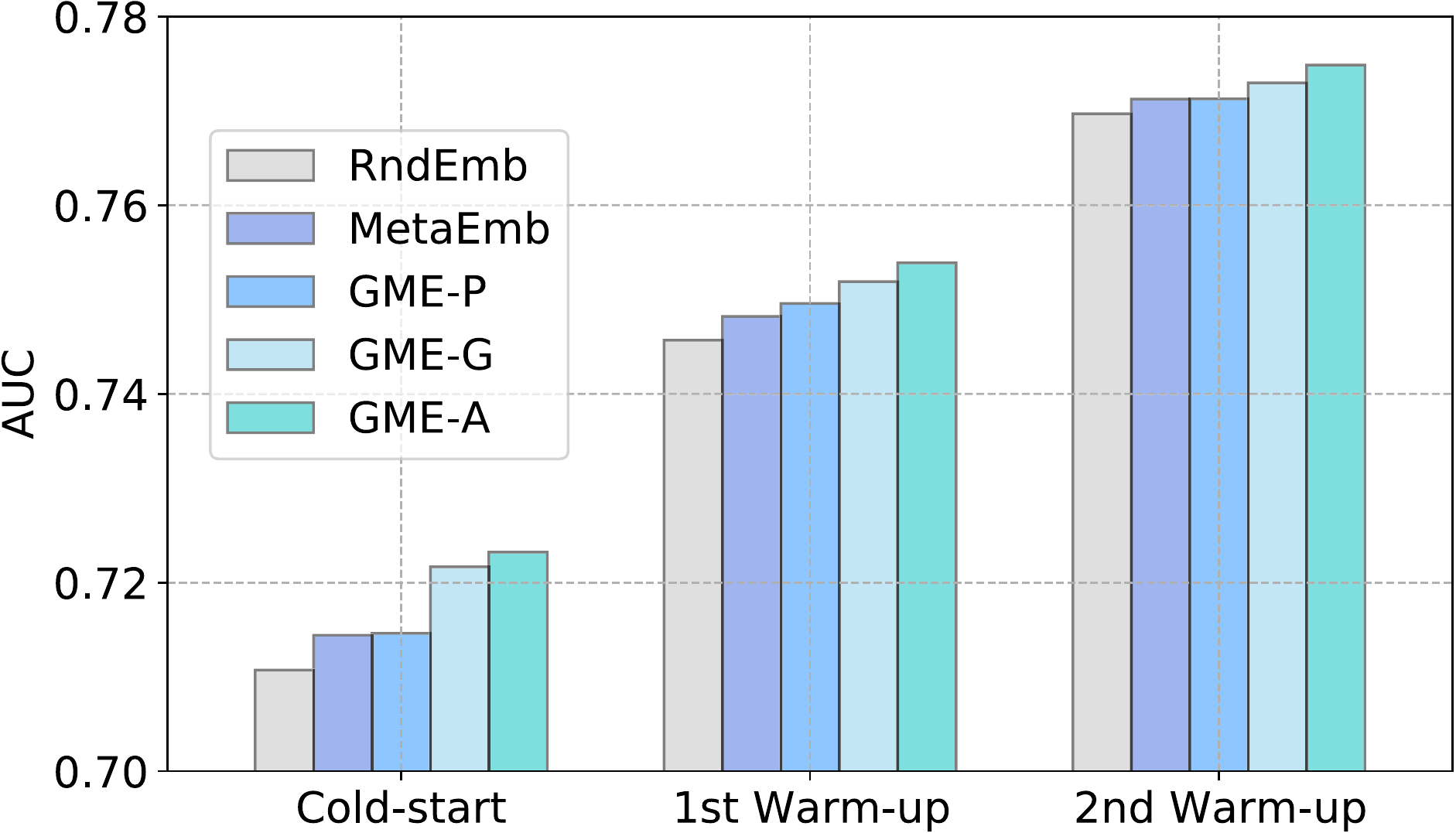}}
\subfigure[Taobao]{\includegraphics[width=0.235\textwidth, trim = 0 0 0 0, clip]{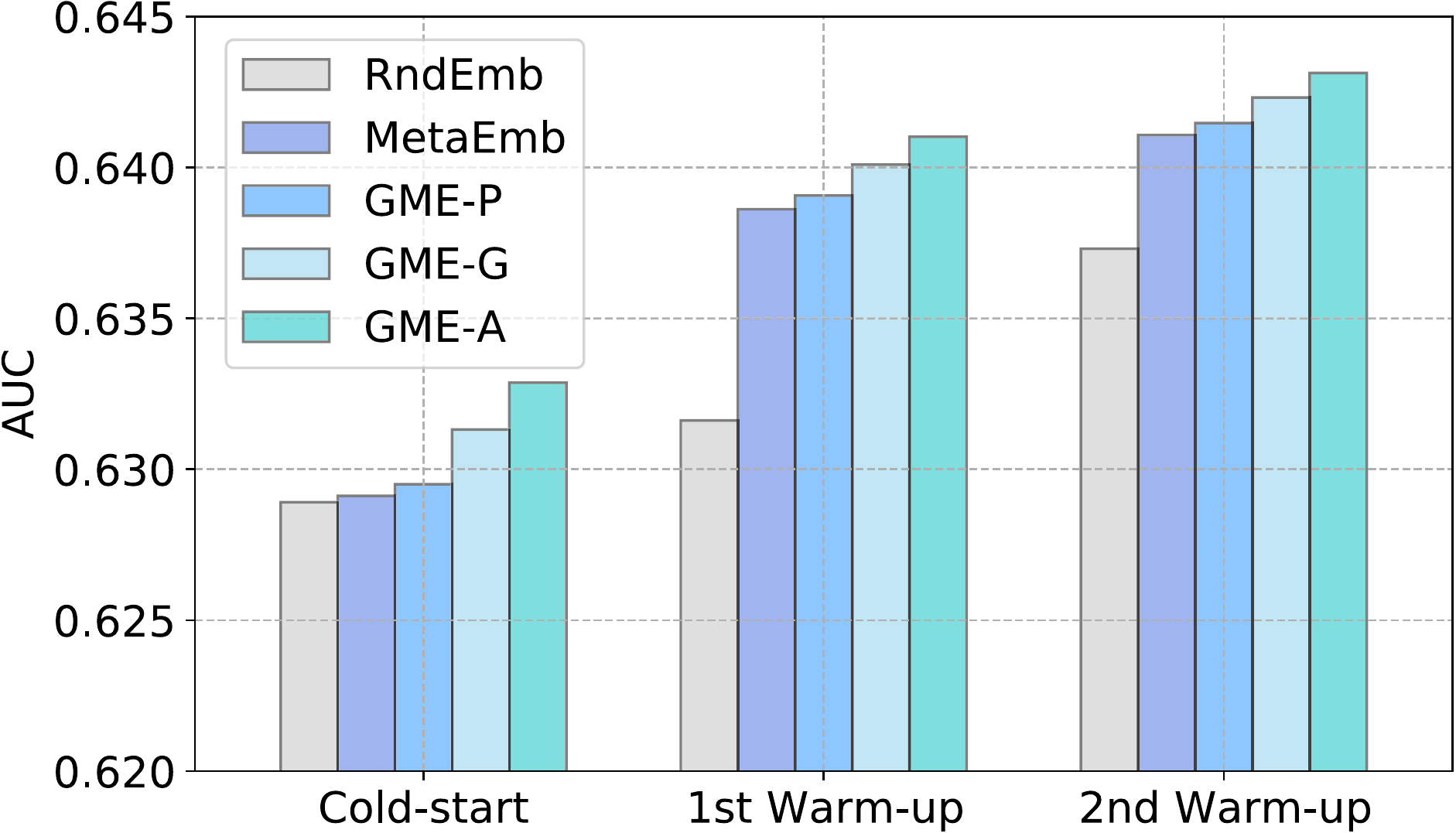}}
\vskip -10pt
\caption{Performance in the warm-up phase. Main prediction model: DNN.}
\vskip -6pt
\label{warm_up}
\end{figure}

Figure \ref{warm_up} plots the performance of various models in the warm-up phase. We perform two rounds of warm-up training.
In the first warm-up training, we provide a small number of training examples (related to new ads) to the main CTR models, but with different initial ID embeddings given by different embedding generation models.
In the second warm-up training, we provide another small number of training examples (related to new ads) to the main CTR models, but based on different ID embeddings learned after the first warm-up training.
It is observed that a model that results in good performance in the cold-start phase generally leads to good performance in the warm-up phase. GME-A not only performs best in the cold-start phase, but also in the two warm-up rounds.

\subsection{Ablation Studies}
\subsubsection{\textbf{Effect of the Scaling Parameter}}
\begin{figure}[!t]
\centering
\subfigure[ML-1M]{\includegraphics[width=0.235\textwidth, trim = 0 0 0 0, clip]{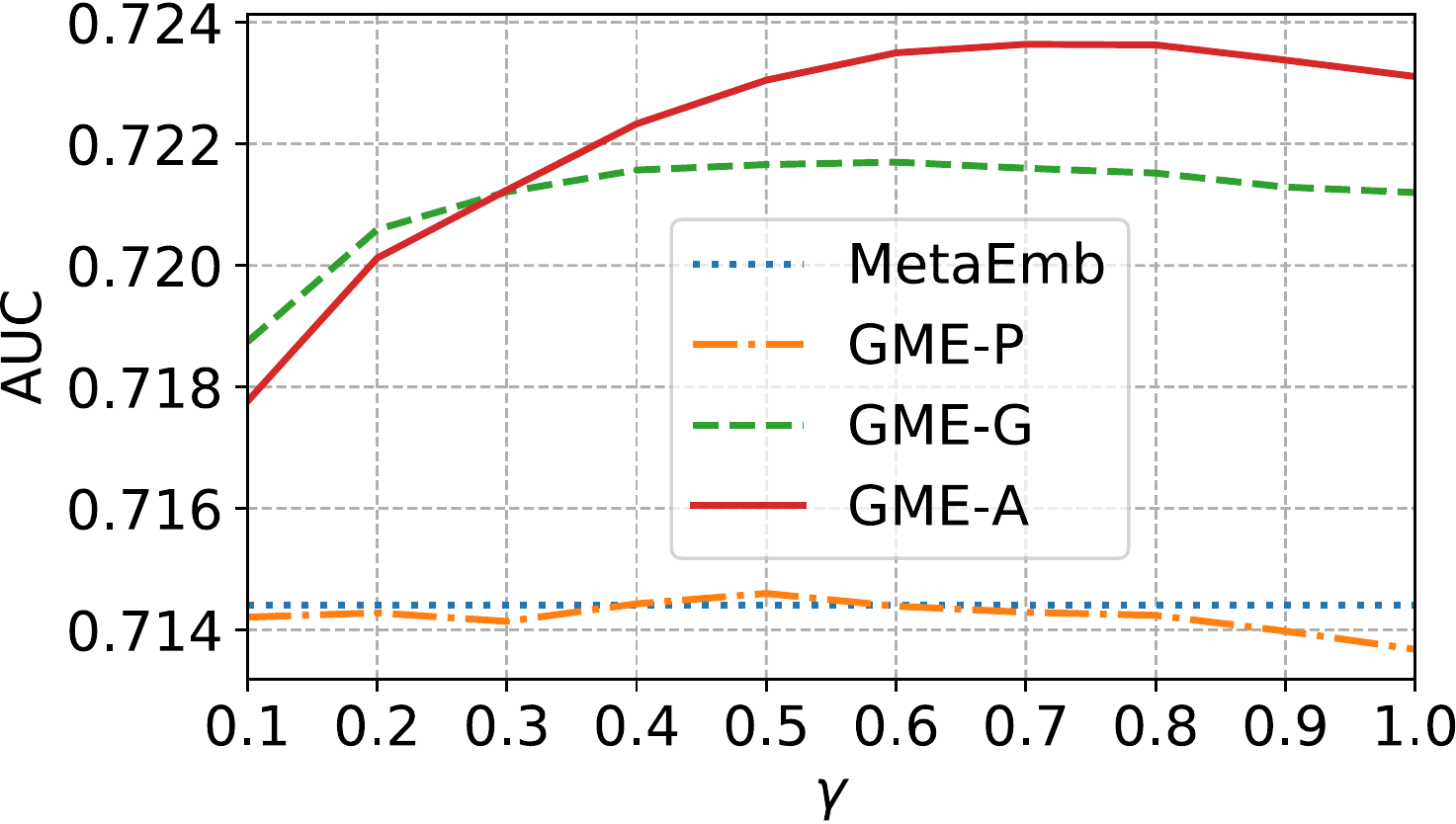}}
\subfigure[Taobao]{\includegraphics[width=0.235\textwidth, trim = 0 0 0 0, clip]{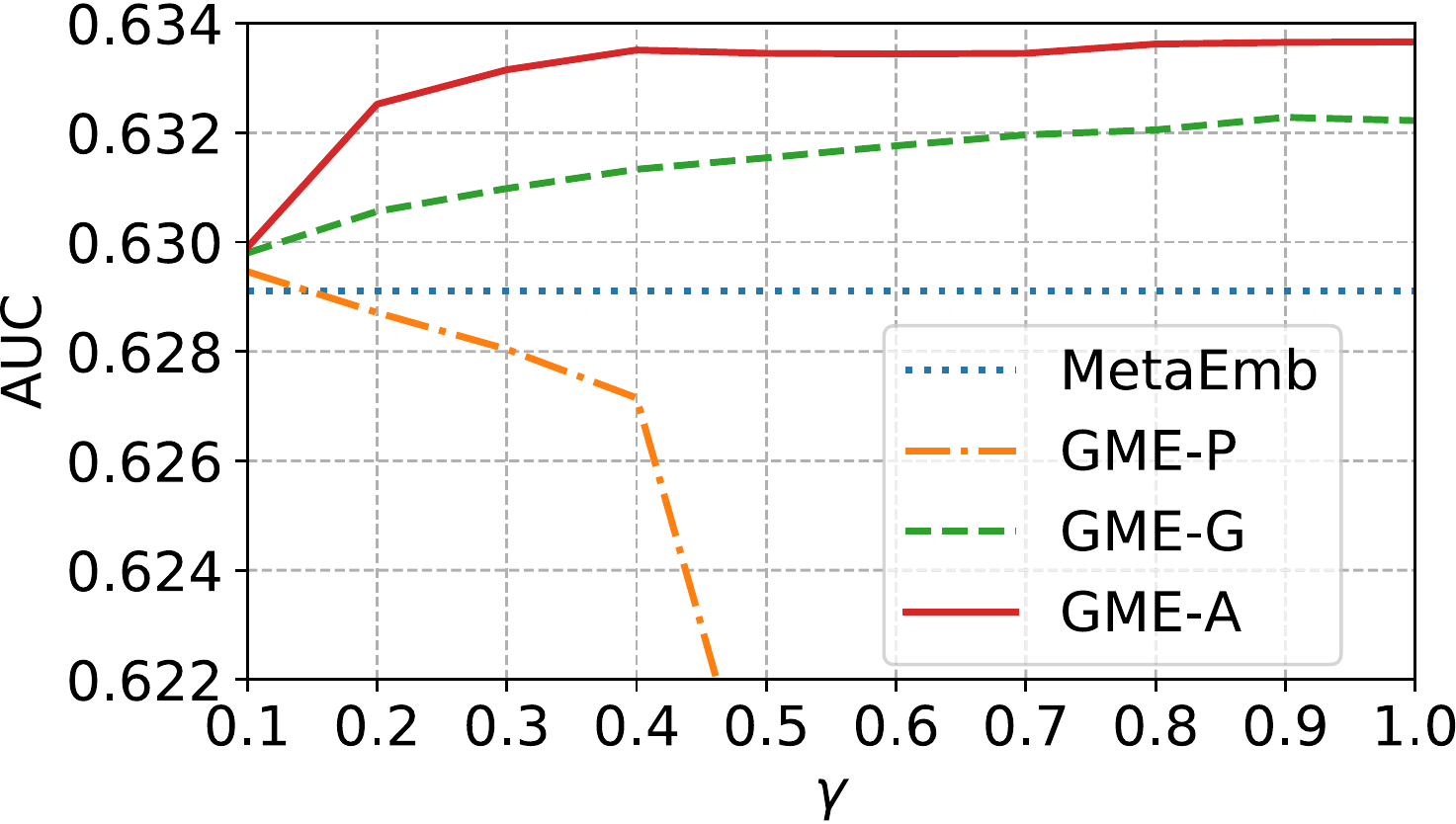}}
\vskip -10pt
\caption{Effect of the scaling parameter. Main prediction model: DNN.}
\vskip -6pt
\label{scaling}
\end{figure}

Figure \ref{scaling} plots the AUC of various models vs. the value of the scaling parameter $\gamma$. On the ML-1M dataset, it is observed that GME-P is relatively insensitive to $\gamma$. Differently, GME-G and GME-A perform much better when $\gamma$ is large. On the Taobao dataset, GME-P performs better when $\gamma$ is small while GME-G and GME-A perform better when $\gamma$ is large. GME-A performs well on a relatively wide range of $\gamma$ values.

\subsubsection{\textbf{Effect of the Number of Neighbors}}
\begin{figure}[!t]
\centering
\subfigure[ML-1M]{\includegraphics[width=0.235\textwidth, trim = 0 0 0 0, clip]{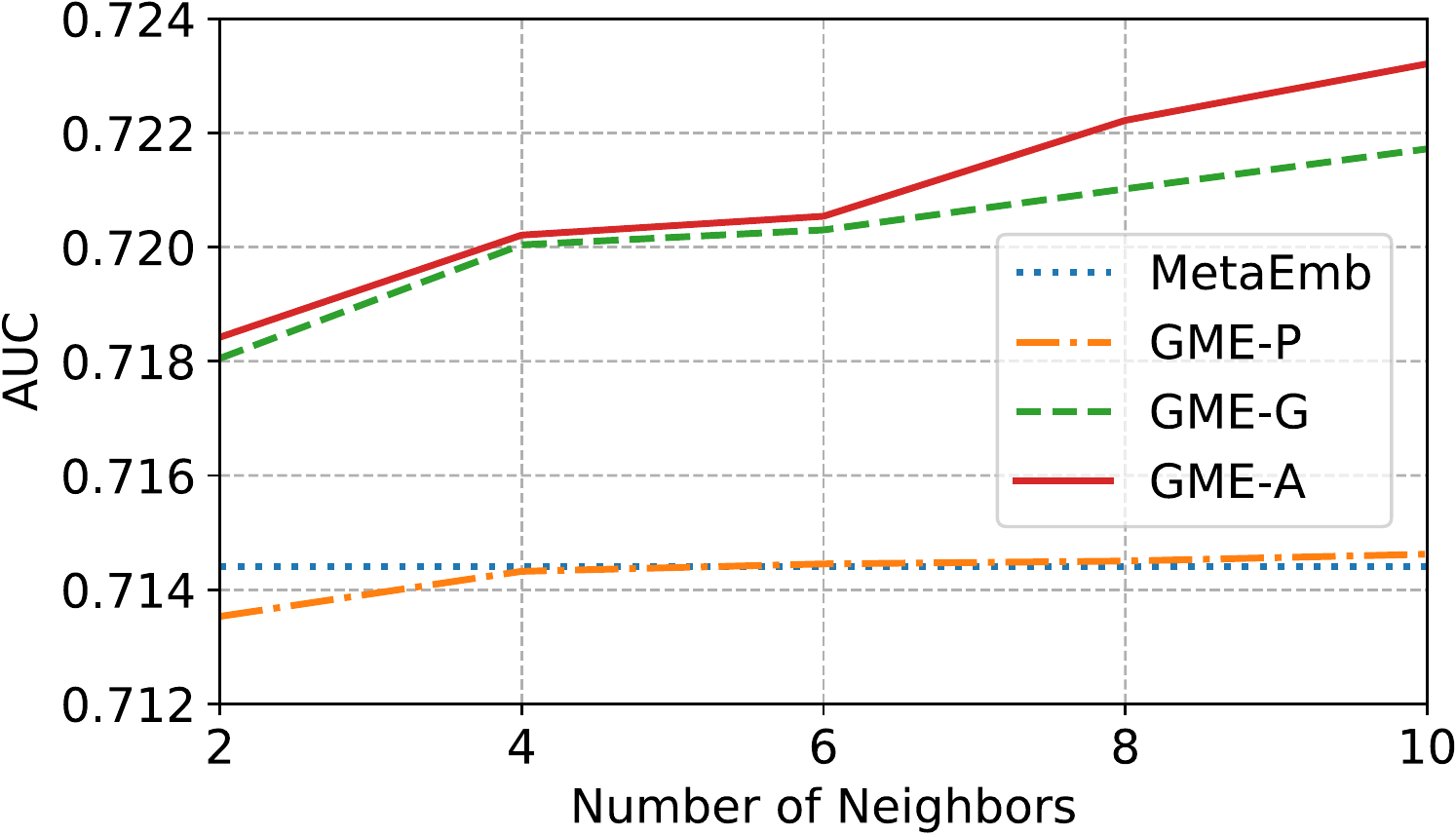}}
\subfigure[Taobao]{\includegraphics[width=0.235\textwidth, trim = 0 0 0 0, clip]{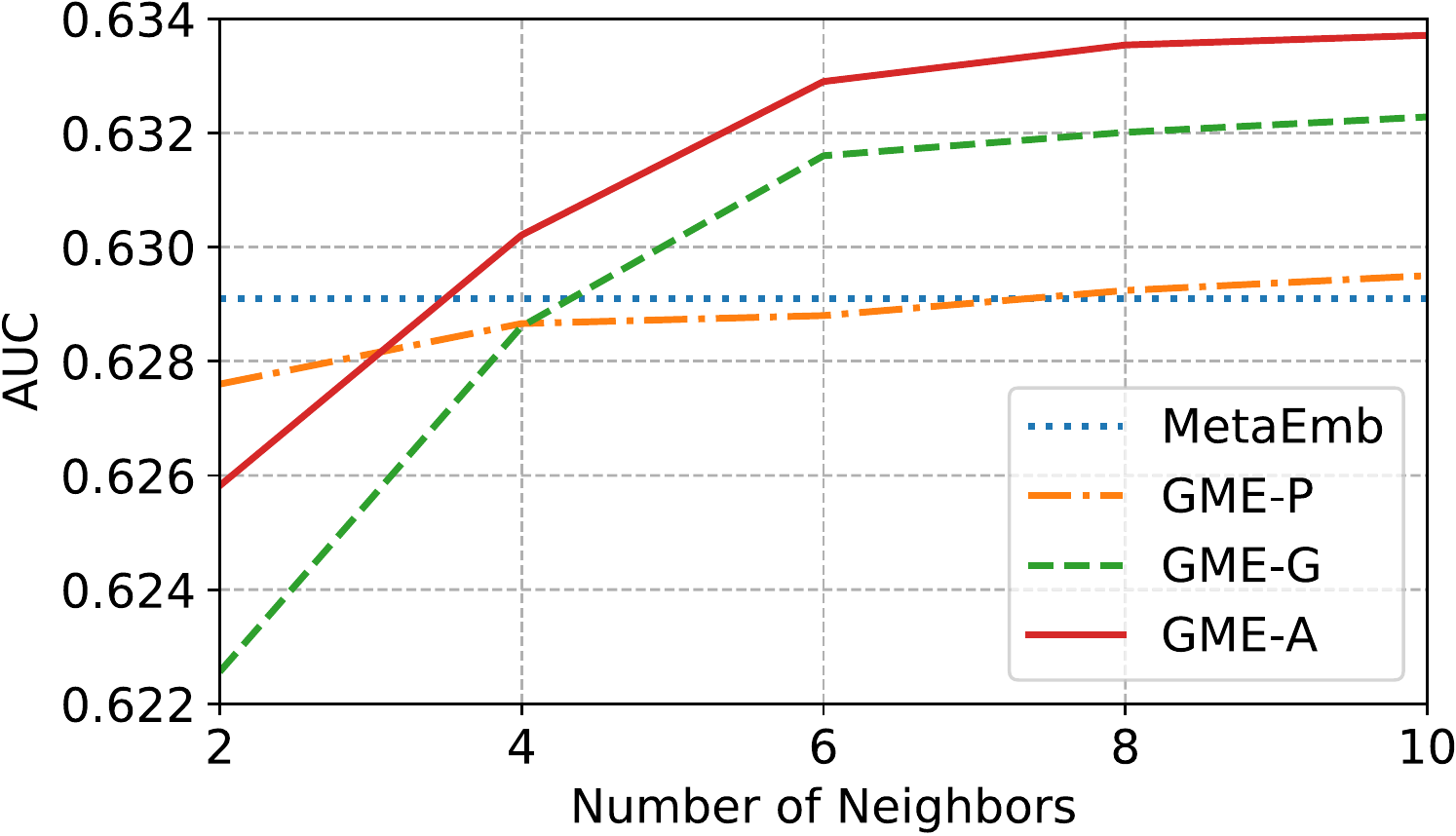}}
\vskip -10pt
\caption{Effect of the number of neighbors. Main prediction model: DNN.}
\vskip -6pt
\label{n_ngb}
\end{figure}

Figure \ref{n_ngb} plots the AUC of various models vs. the number of graph neighbors. It is observed that generally when more neighbors are available, the performance of various GME models also improves. But the performance may become flattened with enough number of neighbors, e.g., the performance of GME-G does not change much when the number of neighbors ranges from 6 to 10 on the Taobao dataset. Moreover, some GME models may not outperform MetaEmb when the number of neighbors is too small (e.g., 2 neighbors on the Taobao dataset).
This is possibly because the neighbors also contain noisy information and it is hard to extract enough useful information from too few neighbors.
Therefore, an enhanced approach to retrieving graph neighbors may lead to further improved performance.

\begin{table}[!t]
\setlength{\tabcolsep}{3pt}
\renewcommand{\arraystretch}{1.2}
\caption{Effect of the GAT. Main prediction model: DNN.}
\vskip -6pt
\label{tab_gat}
\centering
\begin{tabular}{|l|c|c||c|c||c|c|}
\hline
 & \multicolumn{2}{|c||}{\textbf{ML-1M}} & \multicolumn{2}{|c||}{\textbf{Taobao}}  & \multicolumn{2}{|c|}{\textbf{News Feed}} \\
\hline
\textbf{Emb. Model} & AUC & Loss & AUC & Loss & AUC & Loss \\
\hline
GME-P{\textbackslash}GAT & 0.7131 & 0.6447 & 0.6282 & .03208 & 0.7343 & .03636 \\
GME-P & \textbf{0.7146} & \textbf{0.6437} & \textbf{0.6295} & \textbf{.03177} & \textbf{0.7358} & \textbf{.03602} \\
\hline
GME-G{\textbackslash}GAT & 0.7154 & 0.6435 & 0.6301 & .03176 & 0.7355 & .03598 \\
GME-G & \textbf{0.7217} & \textbf{0.6389} & \textbf{0.6323} & \textbf{.03172} & \textbf{0.7371} & \textbf{.03562} \\
\hline
GME-A{\textbackslash}GAT & 0.7156 & 0.6434 & 0.6304 & .03176 & 0.7358 & .03596 \\
GME-A & \textbf{0.7232} & \textbf{0.6368} & \textbf{0.6336} & \textbf{.03168} & \textbf{0.7389} & \textbf{.03553} \\
\hline
\end{tabular}
\vskip -8pt
\end{table}

\subsubsection{\textbf{Effect of the GAT}}
Table \ref{tab_gat} lists the AUC of the GME models with and without the GAT component. When GAT is not used, we aggregates the corresponding representations using average pooling. It is observed that the inclusion of GAT can highly boost the AUC. For example, on the ML-1M dataset, GME-A performs much better than GME-A{\textbackslash}GAT. Moreover, GME-A{\textbackslash}GAT only slightly outperforms GME-G{\textbackslash}GAT. But GME-A largely outperforms GME-G. These results show that GAT can better extract useful information from neighbors than simple average pooling by assigning different importance according to different neighbors' properties. Moreover, applying GAT on ad attributes leads to better performance than applying GAT on generated ID embeddings.

\subsection{Lessons Learned}
We discuss some lessons learned during the experimentation with GME models.

(1) \textbf{Importance of ad IDs.} One would apply the GME models only when the missing of ad IDs impacts the prediction performance significantly. It depends on the property of each specific dataset. In other words, if the exclusion of ad IDs does not degrade the AUC significantly, there is no need to generate better initial embeddings for these IDs.

(2) \textbf{Intrinsic ad attributes.} Ad attributes used to create the ad graph and to generate initial ID embeddings should be intrinsic ad attributes. That is to say, given an ad ID, the associated ad attributes should not change in different samples. Otherwise, we use some changing attributes to generate a fixed ID embedding vector, the model would not be well trained. For example, a specific ad may be displayed at position 1 in one impression and then at position 2 in another impression. Ad position thus can not be used in the aforementioned processes.

(3) \textbf{Positive samples.} When training the ID embedding models with meta learning, there should be some positive samples in most minibatches. One can set the number $M$ larger for datasets with a small proportion of positive samples, perform random sampling multiple times and train the model multiple rounds.

\section{Related Work}
\textbf{CTR prediction.}
The task of CTR prediction in online advertising is to estimate the probability of a user clicking on a specific ad.

As generalized linear models such as Logistic Regression (LR) \cite{richardson2007predicting} and Follow-The-Regularized-Leader (FTRL) \cite{mcmahan2013ad} lack the ability to learn sophisticated feature interactions \cite{chapelle2015simple}, Factorization Machine (FM) \cite{rendle2010factorization, blondel2016higher}, Field-aware FM \cite{juan2016field} and Field-weighted FM \cite{pan2018field} are proposed to address this limitation.

In recent years, deep learning-based models such as Deep Neural Network (DNN) \cite{cheng2016wide}, Product-based Neural Network (PNN) \cite{qu2016product}, Wide\&Deep \cite{cheng2016wide}, DeepFM \cite{guo2017deepfm}, xDeepFM \cite{lian2018xdeepfm} and AutoInt \cite{song2019autoint} are proposed to automatically learn latent feature representations and complicated feature interactions in different manners.
Deep Matching and Prediction (DeepMP) model \cite{ouyang2019representation} combines two subnets to learn more representative feature embeddings for CTR prediction.

Some other models exploit auxiliary information. For example, Deep Interest Network (DIN) \cite{zhou2018deep} and Deep Interest Evolution Network (DIEN) \cite{zhou2019deep} model user interest based on historical click behavior. Xiong et al. \cite{xiong2012relational} and Yin et al. \cite{yin2014exploiting} consider various contextual factors such as ad interaction, ad depth and query diversity. Deep Spatio-Temporal Network (DSTN) \cite{ouyang2019deep} jointly exploits contextual ads, clicked ads and unclicked ads for CTR prediction. Mixed Interest Network (MiNet) \cite{ouyang2020minet} models long- and short-term interests in the news and ads for cross-domain CTR prediction.

However, these models do not specifically address the cold-start problem and they usually have unsatisfactory performance on new ads whose IDs are not seen in the training data.

\textbf{Cold-start recommendation / Cold-start CTR prediction.}
Recommender systems aim to model users' preference on items based on their past interactions. Popular recommendation techniques such as matrix factorization (MF) \cite{koren2009matrix}, neural matrix factorization (NeuMF) \cite{he2017neural} and their families only utilize user IDs and item IDs. Some methods thus propose to use side information for the cold-start scenario, e.g., using user attributes \cite{roy2016latent,seroussi2011personalised,zhang2014addressing} and/or item attributes \cite{saveski2014item,schein2002methods,vartak2017meta,zhang2014addressing}.
However, in the CTR prediction task, side information is already used. The aforementioned CTR prediction models are all feature-rich models, which already take user and ad attributes as input.

Another way to tackle this problem is to actively collect more training data in a short time. For example, \cite{li2010contextual,nguyen2014cold,shah2017practical,tang2015personalized} use contextual-bandit approaches and \cite{golbandi2011adaptive,harpale2008personalized,park2006naive,zhou2011functional} design interviews to collect specific information with active learning. However, these approaches still cannot lead to satisfactory prediction performance before sufficient training data are collected.

We tackle the cold-start CTR prediction problem for new ads from a different perspective, which is to generate desirable initial embeddings for new ad IDs in a meta learning framework, even when the new ads have no training data at all. Along this line, Pan et al. propose the Meta-Embedding model \cite{pan2019warm} by exploiting the associated attributes of the new ad. However, this model only considers the new ad itself, but ignores possibly useful information contained in existing old ads.
Another meta learning-based model MeLU \cite{lee2019melu} is proposed to estimate a new user's preferences with a few consumed items. This model does not apply to our problem and it also considers the target user alone.

\textbf{Meta Learning.}
Meta learning intends to design models that can learn new skills or adapt to new environments rapidly with a few training examples.
It has been successfully applied in various areas such as recommendation \cite{volkovs2017dropoutnet,lee2019melu,lu2020meta}, natural language processing \cite{kiela2018dynamic,xu2018lifelong,chen2018meta} and computer vision \cite{choi2019deep,finn2017model,perez2020incremental}.

There are three common meta learning approaches: 1) metric-based: learn an efficient distance metric, 2) model-based: use (recurrent) networks with external or internal memory, and 3) optimization-based: optimize the model parameters explicitly for fast learning.
The meta learning approach we used to train GMEs is optimization-based, which generalizes Model-Agnostic Meta-Learning (MAML) \cite{finn2017model}. We view the learning of ID embedding of each ad as a task. We use meta learning because the number of unique ads is much smaller than the number of samples and we need fast adaptation.

\section{Conclusion}
In this paper, we address the cold-start CTR prediction problem for new ads whose ID embeddings are not well learned yet.
We propose Graph Meta Embedding (GME) models that can rapidly learn how to generate desirable initial embeddings for new ad IDs based on graph neural networks and meta learning. Unlike previous works that consider the new ad itself, GMEs simultaneously consider two information sources: the new ad and existing old ads. GMEs build a graph to connect new ads and old ads, and adaptively distill useful information from neighboring old ads w.r.t. each given new ad. We propose three specific GMEs from different perspectives. Experimental results show that GMEs can significantly improve the prediction performance in both cold-start and warm-up scenarios over five major deep learning-based CTR prediction models. GME-A which uses neighbor attributes performs best in most cases. In the future, we would consider enhanced approaches to retrieving more informative graph neighbors and alternative ways to distilling more representative information from neighbors.

\bibliographystyle{ACM-Reference-Format}
\balance
\bibliography{ref}


\begin{thebibliography}{58}


\ifx \showCODEN    \undefined \def \showCODEN     #1{\unskip}     \fi
\ifx \showDOI      \undefined \def \showDOI       #1{#1}\fi
\ifx \showISBNx    \undefined \def \showISBNx     #1{\unskip}     \fi
\ifx \showISBNxiii \undefined \def \showISBNxiii  #1{\unskip}     \fi
\ifx \showISSN     \undefined \def \showISSN      #1{\unskip}     \fi
\ifx \showLCCN     \undefined \def \showLCCN      #1{\unskip}     \fi
\ifx \shownote     \undefined \def \shownote      #1{#1}          \fi
\ifx \showarticletitle \undefined \def \showarticletitle #1{#1}   \fi
\ifx \showURL      \undefined \def \showURL       {\relax}        \fi
\providecommand\bibfield[2]{#2}
\providecommand\bibinfo[2]{#2}
\providecommand\natexlab[1]{#1}
\providecommand\showeprint[2][]{arXiv:#2}

\bibitem[\protect\citeauthoryear{Abadi, Barham, Chen, Chen, Davis, Dean, Devin,
  Ghemawat, Irving, Isard, et~al\mbox{.}}{Abadi et~al\mbox{.}}{2016}]%
        {abadi2016tensorflow}
\bibfield{author}{\bibinfo{person}{Mart{\'\i}n Abadi}, \bibinfo{person}{Paul
  Barham}, \bibinfo{person}{Jianmin Chen}, \bibinfo{person}{Zhifeng Chen},
  \bibinfo{person}{Andy Davis}, \bibinfo{person}{Jeffrey Dean},
  \bibinfo{person}{Matthieu Devin}, \bibinfo{person}{Sanjay Ghemawat},
  \bibinfo{person}{Geoffrey Irving}, \bibinfo{person}{Michael Isard},
  {et~al\mbox{.}}} \bibinfo{year}{2016}\natexlab{}.
\newblock \showarticletitle{Tensorflow: A system for large-scale machine
  learning}. In \bibinfo{booktitle}{\emph{OSDI}}. USENIX,
  \bibinfo{pages}{265--283}.
\newblock


\bibitem[\protect\citeauthoryear{Bahdanau, Cho, and Bengio}{Bahdanau
  et~al\mbox{.}}{2015}]%
        {bahdanau2014neural}
\bibfield{author}{\bibinfo{person}{Dzmitry Bahdanau},
  \bibinfo{person}{Kyunghyun Cho}, {and} \bibinfo{person}{Yoshua Bengio}.}
  \bibinfo{year}{2015}\natexlab{}.
\newblock \showarticletitle{Neural machine translation by jointly learning to
  align and translate}. In \bibinfo{booktitle}{\emph{ICLR}}.
\newblock


\bibitem[\protect\citeauthoryear{Blondel, Fujino, Ueda, and Ishihata}{Blondel
  et~al\mbox{.}}{2016}]%
        {blondel2016higher}
\bibfield{author}{\bibinfo{person}{Mathieu Blondel}, \bibinfo{person}{Akinori
  Fujino}, \bibinfo{person}{Naonori Ueda}, {and} \bibinfo{person}{Masakazu
  Ishihata}.} \bibinfo{year}{2016}\natexlab{}.
\newblock \showarticletitle{Higher-order factorization machines}. In
  \bibinfo{booktitle}{\emph{NIPS}}. \bibinfo{pages}{3351--3359}.
\newblock


\bibitem[\protect\citeauthoryear{Chanpuriya and Musco}{Chanpuriya and
  Musco}{2020}]%
        {chanpuriya2020infinitewalk}
\bibfield{author}{\bibinfo{person}{Sudhanshu Chanpuriya} {and}
  \bibinfo{person}{Cameron Musco}.} \bibinfo{year}{2020}\natexlab{}.
\newblock \showarticletitle{Infinitewalk: Deep network embeddings as Laplacian
  embeddings with a nonlinearity}. In \bibinfo{booktitle}{\emph{KDD}}. ACM,
  \bibinfo{pages}{1325--1333}.
\newblock


\bibitem[\protect\citeauthoryear{Chapelle, Manavoglu, and Rosales}{Chapelle
  et~al\mbox{.}}{2015}]%
        {chapelle2015simple}
\bibfield{author}{\bibinfo{person}{Olivier Chapelle}, \bibinfo{person}{Eren
  Manavoglu}, {and} \bibinfo{person}{Romer Rosales}.}
  \bibinfo{year}{2015}\natexlab{}.
\newblock \showarticletitle{Simple and scalable response prediction for display
  advertising}.
\newblock \bibinfo{journal}{\emph{ACM TIST}} \bibinfo{volume}{5},
  \bibinfo{number}{4} (\bibinfo{year}{2015}), \bibinfo{pages}{61}.
\newblock


\bibitem[\protect\citeauthoryear{Chen, Qiu, Liu, and Huang}{Chen
  et~al\mbox{.}}{2018}]%
        {chen2018meta}
\bibfield{author}{\bibinfo{person}{Junkun Chen}, \bibinfo{person}{Xipeng Qiu},
  \bibinfo{person}{Pengfei Liu}, {and} \bibinfo{person}{Xuanjing Huang}.}
  \bibinfo{year}{2018}\natexlab{}.
\newblock \showarticletitle{Meta multi-task learning for sequence modeling}. In
  \bibinfo{booktitle}{\emph{AAAI}}, Vol.~\bibinfo{volume}{32}.
\newblock


\bibitem[\protect\citeauthoryear{Cheng, Koc, Harmsen, Shaked, Chandra, Aradhye,
  Anderson, Corrado, Chai, Ispir, et~al\mbox{.}}{Cheng et~al\mbox{.}}{2016}]%
        {cheng2016wide}
\bibfield{author}{\bibinfo{person}{Heng-Tze Cheng}, \bibinfo{person}{Levent
  Koc}, \bibinfo{person}{Jeremiah Harmsen}, \bibinfo{person}{Tal Shaked},
  \bibinfo{person}{Tushar Chandra}, \bibinfo{person}{Hrishi Aradhye},
  \bibinfo{person}{Glen Anderson}, \bibinfo{person}{Greg Corrado},
  \bibinfo{person}{Wei Chai}, \bibinfo{person}{Mustafa Ispir}, {et~al\mbox{.}}}
  \bibinfo{year}{2016}\natexlab{}.
\newblock \showarticletitle{Wide \& deep learning for recommender systems}. In
  \bibinfo{booktitle}{\emph{DLRS}}. ACM, \bibinfo{pages}{7--10}.
\newblock


\bibitem[\protect\citeauthoryear{Choi, Kwon, and Lee}{Choi
  et~al\mbox{.}}{2019}]%
        {choi2019deep}
\bibfield{author}{\bibinfo{person}{Janghoon Choi}, \bibinfo{person}{Junseok
  Kwon}, {and} \bibinfo{person}{Kyoung~Mu Lee}.}
  \bibinfo{year}{2019}\natexlab{}.
\newblock \showarticletitle{Deep meta learning for real-time target-aware
  visual tracking}. In \bibinfo{booktitle}{\emph{CVPR}}. IEEE,
  \bibinfo{pages}{911--920}.
\newblock


\bibitem[\protect\citeauthoryear{Finn, Abbeel, and Levine}{Finn
  et~al\mbox{.}}{2017}]%
        {finn2017model}
\bibfield{author}{\bibinfo{person}{Chelsea Finn}, \bibinfo{person}{Pieter
  Abbeel}, {and} \bibinfo{person}{Sergey Levine}.}
  \bibinfo{year}{2017}\natexlab{}.
\newblock \showarticletitle{Model-Agnostic Meta-Learning for Fast Adaptation of
  Deep Networks}. In \bibinfo{booktitle}{\emph{ICML}}.
  \bibinfo{pages}{1126--1135}.
\newblock


\bibitem[\protect\citeauthoryear{Golbandi, Koren, and Lempel}{Golbandi
  et~al\mbox{.}}{2011}]%
        {golbandi2011adaptive}
\bibfield{author}{\bibinfo{person}{Nadav Golbandi}, \bibinfo{person}{Yehuda
  Koren}, {and} \bibinfo{person}{Ronny Lempel}.}
  \bibinfo{year}{2011}\natexlab{}.
\newblock \showarticletitle{Adaptive bootstrapping of recommender systems using
  decision trees}. In \bibinfo{booktitle}{\emph{WSDM}}. ACM,
  \bibinfo{pages}{595--604}.
\newblock


\bibitem[\protect\citeauthoryear{Guo, Tang, Ye, Li, and He}{Guo
  et~al\mbox{.}}{2017}]%
        {guo2017deepfm}
\bibfield{author}{\bibinfo{person}{Huifeng Guo}, \bibinfo{person}{Ruiming
  Tang}, \bibinfo{person}{Yunming Ye}, \bibinfo{person}{Zhenguo Li}, {and}
  \bibinfo{person}{Xiuqiang He}.} \bibinfo{year}{2017}\natexlab{}.
\newblock \showarticletitle{Deepfm: a factorization-machine based neural
  network for ctr prediction}. In \bibinfo{booktitle}{\emph{IJCAI}}.
  \bibinfo{pages}{1725--1731}.
\newblock


\bibitem[\protect\citeauthoryear{Harpale and Yang}{Harpale and Yang}{2008}]%
        {harpale2008personalized}
\bibfield{author}{\bibinfo{person}{Abhay~S Harpale} {and}
  \bibinfo{person}{Yiming Yang}.} \bibinfo{year}{2008}\natexlab{}.
\newblock \showarticletitle{Personalized active learning for collaborative
  filtering}. In \bibinfo{booktitle}{\emph{SIGIR}}. ACM,
  \bibinfo{pages}{91--98}.
\newblock


\bibitem[\protect\citeauthoryear{He, Liao, Zhang, Nie, Hu, and Chua}{He
  et~al\mbox{.}}{2017}]%
        {he2017neural}
\bibfield{author}{\bibinfo{person}{Xiangnan He}, \bibinfo{person}{Lizi Liao},
  \bibinfo{person}{Hanwang Zhang}, \bibinfo{person}{Liqiang Nie},
  \bibinfo{person}{Xia Hu}, {and} \bibinfo{person}{Tat-Seng Chua}.}
  \bibinfo{year}{2017}\natexlab{}.
\newblock \showarticletitle{Neural collaborative filtering}. In
  \bibinfo{booktitle}{\emph{WWW}}. \bibinfo{pages}{173--182}.
\newblock


\bibitem[\protect\citeauthoryear{He, Pan, Jin, Xu, Liu, Xu, Shi, Atallah,
  Herbrich, Bowers, et~al\mbox{.}}{He et~al\mbox{.}}{2014}]%
        {he2014practical}
\bibfield{author}{\bibinfo{person}{Xinran He}, \bibinfo{person}{Junfeng Pan},
  \bibinfo{person}{Ou Jin}, \bibinfo{person}{Tianbing Xu}, \bibinfo{person}{Bo
  Liu}, \bibinfo{person}{Tao Xu}, \bibinfo{person}{Yanxin Shi},
  \bibinfo{person}{Antoine Atallah}, \bibinfo{person}{Ralf Herbrich},
  \bibinfo{person}{Stuart Bowers}, {et~al\mbox{.}}}
  \bibinfo{year}{2014}\natexlab{}.
\newblock \showarticletitle{Practical lessons from predicting clicks on ads at
  facebook}. In \bibinfo{booktitle}{\emph{ADKDD}}. ACM, \bibinfo{pages}{1--9}.
\newblock


\bibitem[\protect\citeauthoryear{Juan, Zhuang, Chin, and Lin}{Juan
  et~al\mbox{.}}{2016}]%
        {juan2016field}
\bibfield{author}{\bibinfo{person}{Yuchin Juan}, \bibinfo{person}{Yong Zhuang},
  \bibinfo{person}{Wei-Sheng Chin}, {and} \bibinfo{person}{Chih-Jen Lin}.}
  \bibinfo{year}{2016}\natexlab{}.
\newblock \showarticletitle{Field-aware factorization machines for CTR
  prediction}. In \bibinfo{booktitle}{\emph{RecSys}}. ACM,
  \bibinfo{pages}{43--50}.
\newblock


\bibitem[\protect\citeauthoryear{Kiela, Wang, and Cho}{Kiela
  et~al\mbox{.}}{2018}]%
        {kiela2018dynamic}
\bibfield{author}{\bibinfo{person}{Douwe Kiela}, \bibinfo{person}{Changhan
  Wang}, {and} \bibinfo{person}{Kyunghyun Cho}.}
  \bibinfo{year}{2018}\natexlab{}.
\newblock \showarticletitle{Dynamic Meta-Embeddings for Improved Sentence
  Representations}. In \bibinfo{booktitle}{\emph{EMNLP}}.
  \bibinfo{pages}{1466--1477}.
\newblock


\bibitem[\protect\citeauthoryear{Kingma and Ba}{Kingma and Ba}{2014}]%
        {kingma2014adam}
\bibfield{author}{\bibinfo{person}{Diederik~P Kingma} {and}
  \bibinfo{person}{Jimmy Ba}.} \bibinfo{year}{2014}\natexlab{}.
\newblock \showarticletitle{Adam: A method for stochastic optimization}.
\newblock \bibinfo{journal}{\emph{arXiv preprint arXiv:1412.6980}}
  (\bibinfo{year}{2014}).
\newblock


\bibitem[\protect\citeauthoryear{Koren, Bell, and Volinsky}{Koren
  et~al\mbox{.}}{2009}]%
        {koren2009matrix}
\bibfield{author}{\bibinfo{person}{Yehuda Koren}, \bibinfo{person}{Robert
  Bell}, {and} \bibinfo{person}{Chris Volinsky}.}
  \bibinfo{year}{2009}\natexlab{}.
\newblock \showarticletitle{Matrix factorization techniques for recommender
  systems}.
\newblock \bibinfo{journal}{\emph{Computer}} \bibinfo{volume}{42},
  \bibinfo{number}{8} (\bibinfo{year}{2009}), \bibinfo{pages}{30--37}.
\newblock


\bibitem[\protect\citeauthoryear{Lee, Im, Jang, Cho, and Chung}{Lee
  et~al\mbox{.}}{2019}]%
        {lee2019melu}
\bibfield{author}{\bibinfo{person}{Hoyeop Lee}, \bibinfo{person}{Jinbae Im},
  \bibinfo{person}{Seongwon Jang}, \bibinfo{person}{Hyunsouk Cho}, {and}
  \bibinfo{person}{Sehee Chung}.} \bibinfo{year}{2019}\natexlab{}.
\newblock \showarticletitle{MeLU: meta-learned user preference estimator for
  cold-start recommendation}. In \bibinfo{booktitle}{\emph{KDD}}.
  \bibinfo{pages}{1073--1082}.
\newblock


\bibitem[\protect\citeauthoryear{Li, Chu, Langford, and Schapire}{Li
  et~al\mbox{.}}{2010}]%
        {li2010contextual}
\bibfield{author}{\bibinfo{person}{Lihong Li}, \bibinfo{person}{Wei Chu},
  \bibinfo{person}{John Langford}, {and} \bibinfo{person}{Robert~E Schapire}.}
  \bibinfo{year}{2010}\natexlab{}.
\newblock \showarticletitle{A contextual-bandit approach to personalized news
  article recommendation}. In \bibinfo{booktitle}{\emph{WWW}}. ACM,
  \bibinfo{pages}{661--670}.
\newblock


\bibitem[\protect\citeauthoryear{Lian, Zhou, Zhang, Chen, Xie, and Sun}{Lian
  et~al\mbox{.}}{2018}]%
        {lian2018xdeepfm}
\bibfield{author}{\bibinfo{person}{Jianxun Lian}, \bibinfo{person}{Xiaohuan
  Zhou}, \bibinfo{person}{Fuzheng Zhang}, \bibinfo{person}{Zhongxia Chen},
  \bibinfo{person}{Xing Xie}, {and} \bibinfo{person}{Guangzhong Sun}.}
  \bibinfo{year}{2018}\natexlab{}.
\newblock \showarticletitle{xDeepFM: Combining explicit and implicit feature
  interactions for recommender systems}. In \bibinfo{booktitle}{\emph{KDD}}.
  ACM, \bibinfo{pages}{1754--1763}.
\newblock


\bibitem[\protect\citeauthoryear{Lu, Fang, and Shi}{Lu et~al\mbox{.}}{2020}]%
        {lu2020meta}
\bibfield{author}{\bibinfo{person}{Yuanfu Lu}, \bibinfo{person}{Yuan Fang},
  {and} \bibinfo{person}{Chuan Shi}.} \bibinfo{year}{2020}\natexlab{}.
\newblock \showarticletitle{Meta-learning on heterogeneous information networks
  for cold-start recommendation}. In \bibinfo{booktitle}{\emph{KDD}}. ACM,
  \bibinfo{pages}{1563--1573}.
\newblock


\bibitem[\protect\citeauthoryear{Ma, Zhao, Huang, Wang, Hu, Zhu, and Gai}{Ma
  et~al\mbox{.}}{2018}]%
        {ma2018entire}
\bibfield{author}{\bibinfo{person}{Xiao Ma}, \bibinfo{person}{Liqin Zhao},
  \bibinfo{person}{Guan Huang}, \bibinfo{person}{Zhi Wang},
  \bibinfo{person}{Zelin Hu}, \bibinfo{person}{Xiaoqiang Zhu}, {and}
  \bibinfo{person}{Kun Gai}.} \bibinfo{year}{2018}\natexlab{}.
\newblock \showarticletitle{Entire space multi-task model: An effective
  approach for estimating post-click conversion rate}. In
  \bibinfo{booktitle}{\emph{SIGIR}}. ACM, \bibinfo{pages}{1137--1140}.
\newblock


\bibitem[\protect\citeauthoryear{Maas, Hannun, and Ng}{Maas
  et~al\mbox{.}}{2013}]%
        {maas2013rectifier}
\bibfield{author}{\bibinfo{person}{Andrew~L Maas}, \bibinfo{person}{Awni~Y
  Hannun}, {and} \bibinfo{person}{Andrew~Y Ng}.}
  \bibinfo{year}{2013}\natexlab{}.
\newblock \showarticletitle{Rectifier nonlinearities improve neural network
  acoustic models}. In \bibinfo{booktitle}{\emph{ICML}},
  Vol.~\bibinfo{volume}{30}. \bibinfo{pages}{3}.
\newblock


\bibitem[\protect\citeauthoryear{McMahan, Holt, Sculley, Young, Ebner, Grady,
  Nie, Phillips, Davydov, Golovin, et~al\mbox{.}}{McMahan
  et~al\mbox{.}}{2013}]%
        {mcmahan2013ad}
\bibfield{author}{\bibinfo{person}{H~Brendan McMahan}, \bibinfo{person}{Gary
  Holt}, \bibinfo{person}{David Sculley}, \bibinfo{person}{Michael Young},
  \bibinfo{person}{Dietmar Ebner}, \bibinfo{person}{Julian Grady},
  \bibinfo{person}{Lan Nie}, \bibinfo{person}{Todd Phillips},
  \bibinfo{person}{Eugene Davydov}, \bibinfo{person}{Daniel Golovin},
  {et~al\mbox{.}}} \bibinfo{year}{2013}\natexlab{}.
\newblock \showarticletitle{Ad click prediction: a view from the trenches}. In
  \bibinfo{booktitle}{\emph{KDD}}. ACM, \bibinfo{pages}{1222--1230}.
\newblock


\bibitem[\protect\citeauthoryear{Mikolov, Sutskever, Chen, Corrado, and
  Dean}{Mikolov et~al\mbox{.}}{2013}]%
        {mikolov2013distributed}
\bibfield{author}{\bibinfo{person}{Tomas Mikolov}, \bibinfo{person}{Ilya
  Sutskever}, \bibinfo{person}{Kai Chen}, \bibinfo{person}{Greg~S Corrado},
  {and} \bibinfo{person}{Jeff Dean}.} \bibinfo{year}{2013}\natexlab{}.
\newblock \showarticletitle{Distributed representations of words and phrases
  and their compositionality}. In \bibinfo{booktitle}{\emph{NIPS}}.
  \bibinfo{pages}{3111--3119}.
\newblock


\bibitem[\protect\citeauthoryear{Nguyen, Mary, and Preux}{Nguyen
  et~al\mbox{.}}{2014}]%
        {nguyen2014cold}
\bibfield{author}{\bibinfo{person}{Hai~Thanh Nguyen},
  \bibinfo{person}{J{\'e}r{\'e}mie Mary}, {and} \bibinfo{person}{Philippe
  Preux}.} \bibinfo{year}{2014}\natexlab{}.
\newblock \showarticletitle{Cold-start problems in recommendation systems via
  contextual-bandit algorithms}.
\newblock \bibinfo{journal}{\emph{arXiv preprint arXiv:1405.7544}}
  (\bibinfo{year}{2014}).
\newblock


\bibitem[\protect\citeauthoryear{Ouyang, Zhang, Li, Zou, Xing, Liu, and
  Du}{Ouyang et~al\mbox{.}}{2019a}]%
        {ouyang2019deep}
\bibfield{author}{\bibinfo{person}{Wentao Ouyang}, \bibinfo{person}{Xiuwu
  Zhang}, \bibinfo{person}{Li Li}, \bibinfo{person}{Heng Zou},
  \bibinfo{person}{Xin Xing}, \bibinfo{person}{Zhaojie Liu}, {and}
  \bibinfo{person}{Yanlong Du}.} \bibinfo{year}{2019}\natexlab{a}.
\newblock \showarticletitle{Deep spatio-temporal neural networks for
  click-through rate prediction}. In \bibinfo{booktitle}{\emph{KDD}}. ACM,
  \bibinfo{pages}{2078--2086}.
\newblock


\bibitem[\protect\citeauthoryear{Ouyang, Zhang, Ren, Li, Liu, and Du}{Ouyang
  et~al\mbox{.}}{2019b}]%
        {ouyang2019click}
\bibfield{author}{\bibinfo{person}{Wentao Ouyang}, \bibinfo{person}{Xiuwu
  Zhang}, \bibinfo{person}{Shukui Ren}, \bibinfo{person}{Li Li},
  \bibinfo{person}{Zhaojie Liu}, {and} \bibinfo{person}{Yanlong Du}.}
  \bibinfo{year}{2019}\natexlab{b}.
\newblock \showarticletitle{Click-through rate prediction with the user memory
  network}. In \bibinfo{booktitle}{\emph{DLP-KDD}}. \bibinfo{pages}{1--4}.
\newblock


\bibitem[\protect\citeauthoryear{Ouyang, Zhang, Ren, Qi, Liu, and Du}{Ouyang
  et~al\mbox{.}}{2019c}]%
        {ouyang2019representation}
\bibfield{author}{\bibinfo{person}{Wentao Ouyang}, \bibinfo{person}{Xiuwu
  Zhang}, \bibinfo{person}{Shukui Ren}, \bibinfo{person}{Chao Qi},
  \bibinfo{person}{Zhaojie Liu}, {and} \bibinfo{person}{Yanlong Du}.}
  \bibinfo{year}{2019}\natexlab{c}.
\newblock \showarticletitle{Representation Learning-Assisted Click-Through Rate
  Prediction}. In \bibinfo{booktitle}{\emph{IJCAI}}.
  \bibinfo{pages}{4561--4567}.
\newblock


\bibitem[\protect\citeauthoryear{Ouyang, Zhang, Zhao, Luo, Zhang, Zou, Liu, and
  Du}{Ouyang et~al\mbox{.}}{2020}]%
        {ouyang2020minet}
\bibfield{author}{\bibinfo{person}{Wentao Ouyang}, \bibinfo{person}{Xiuwu
  Zhang}, \bibinfo{person}{Lei Zhao}, \bibinfo{person}{Jinmei Luo},
  \bibinfo{person}{Yu Zhang}, \bibinfo{person}{Heng Zou},
  \bibinfo{person}{Zhaojie Liu}, {and} \bibinfo{person}{Yanlong Du}.}
  \bibinfo{year}{2020}\natexlab{}.
\newblock \showarticletitle{MiNet: Mixed Interest Network for Cross-Domain
  Click-Through Rate Prediction}. In \bibinfo{booktitle}{\emph{CIKM}}. ACM,
  \bibinfo{pages}{2669--2676}.
\newblock


\bibitem[\protect\citeauthoryear{Pan, Li, Ao, Tang, and He}{Pan
  et~al\mbox{.}}{2019}]%
        {pan2019warm}
\bibfield{author}{\bibinfo{person}{Feiyang Pan}, \bibinfo{person}{Shuokai Li},
  \bibinfo{person}{Xiang Ao}, \bibinfo{person}{Pingzhong Tang}, {and}
  \bibinfo{person}{Qing He}.} \bibinfo{year}{2019}\natexlab{}.
\newblock \showarticletitle{Warm up cold-start advertisements: Improving ctr
  predictions via learning to learn id embeddings}. In
  \bibinfo{booktitle}{\emph{SIGIR}}. ACM, \bibinfo{pages}{695--704}.
\newblock


\bibitem[\protect\citeauthoryear{Pan, Xu, Ruiz, Zhao, Pan, Sun, and Lu}{Pan
  et~al\mbox{.}}{2018}]%
        {pan2018field}
\bibfield{author}{\bibinfo{person}{Junwei Pan}, \bibinfo{person}{Jian Xu},
  \bibinfo{person}{Alfonso~Lobos Ruiz}, \bibinfo{person}{Wenliang Zhao},
  \bibinfo{person}{Shengjun Pan}, \bibinfo{person}{Yu Sun}, {and}
  \bibinfo{person}{Quan Lu}.} \bibinfo{year}{2018}\natexlab{}.
\newblock \showarticletitle{Field-weighted Factorization Machines for
  Click-Through Rate Prediction in Display Advertising}. In
  \bibinfo{booktitle}{\emph{WWW}}. IW3C2, \bibinfo{pages}{1349--1357}.
\newblock


\bibitem[\protect\citeauthoryear{Park, Pennock, Madani, Good, and DeCoste}{Park
  et~al\mbox{.}}{2006}]%
        {park2006naive}
\bibfield{author}{\bibinfo{person}{Seung-Taek Park}, \bibinfo{person}{David
  Pennock}, \bibinfo{person}{Omid Madani}, \bibinfo{person}{Nathan Good}, {and}
  \bibinfo{person}{Dennis DeCoste}.} \bibinfo{year}{2006}\natexlab{}.
\newblock \showarticletitle{Naive filterbots for robust cold-start
  recommendations}. In \bibinfo{booktitle}{\emph{KDD}}. ACM,
  \bibinfo{pages}{699--705}.
\newblock


\bibitem[\protect\citeauthoryear{Perez-Rua, Zhu, Hospedales, and
  Xiang}{Perez-Rua et~al\mbox{.}}{2020}]%
        {perez2020incremental}
\bibfield{author}{\bibinfo{person}{Juan-Manuel Perez-Rua},
  \bibinfo{person}{Xiatian Zhu}, \bibinfo{person}{Timothy~M Hospedales}, {and}
  \bibinfo{person}{Tao Xiang}.} \bibinfo{year}{2020}\natexlab{}.
\newblock \showarticletitle{Incremental few-shot object detection}. In
  \bibinfo{booktitle}{\emph{CVPR}}. IEEE, \bibinfo{pages}{13846--13855}.
\newblock


\bibitem[\protect\citeauthoryear{Qin, Zhang, Wu, Jin, Fang, and Yu}{Qin
  et~al\mbox{.}}{2020}]%
        {qin2020user}
\bibfield{author}{\bibinfo{person}{Jiarui Qin}, \bibinfo{person}{Weinan Zhang},
  \bibinfo{person}{Xin Wu}, \bibinfo{person}{Jiarui Jin},
  \bibinfo{person}{Yuchen Fang}, {and} \bibinfo{person}{Yong Yu}.}
  \bibinfo{year}{2020}\natexlab{}.
\newblock \showarticletitle{User Behavior Retrieval for Click-Through Rate
  Prediction}. In \bibinfo{booktitle}{\emph{SIGIR}}. ACM,
  \bibinfo{pages}{2347--2356}.
\newblock


\bibitem[\protect\citeauthoryear{Qu, Cai, Ren, Zhang, Yu, Wen, and Wang}{Qu
  et~al\mbox{.}}{2016}]%
        {qu2016product}
\bibfield{author}{\bibinfo{person}{Yanru Qu}, \bibinfo{person}{Han Cai},
  \bibinfo{person}{Kan Ren}, \bibinfo{person}{Weinan Zhang},
  \bibinfo{person}{Yong Yu}, \bibinfo{person}{Ying Wen}, {and}
  \bibinfo{person}{Jun Wang}.} \bibinfo{year}{2016}\natexlab{}.
\newblock \showarticletitle{Product-based neural networks for user response
  prediction}. In \bibinfo{booktitle}{\emph{ICDM}}. IEEE,
  \bibinfo{pages}{1149--1154}.
\newblock


\bibitem[\protect\citeauthoryear{Rendle}{Rendle}{2010}]%
        {rendle2010factorization}
\bibfield{author}{\bibinfo{person}{Steffen Rendle}.}
  \bibinfo{year}{2010}\natexlab{}.
\newblock \showarticletitle{Factorization machines}. In
  \bibinfo{booktitle}{\emph{ICDM}}. IEEE, \bibinfo{pages}{995--1000}.
\newblock


\bibitem[\protect\citeauthoryear{Richardson, Dominowska, and Ragno}{Richardson
  et~al\mbox{.}}{2007}]%
        {richardson2007predicting}
\bibfield{author}{\bibinfo{person}{Matthew Richardson}, \bibinfo{person}{Ewa
  Dominowska}, {and} \bibinfo{person}{Robert Ragno}.}
  \bibinfo{year}{2007}\natexlab{}.
\newblock \showarticletitle{Predicting clicks: estimating the click-through
  rate for new ads}. In \bibinfo{booktitle}{\emph{WWW}}. IW3C2,
  \bibinfo{pages}{521--530}.
\newblock


\bibitem[\protect\citeauthoryear{Roy and Guntuku}{Roy and Guntuku}{2016}]%
        {roy2016latent}
\bibfield{author}{\bibinfo{person}{Sujoy Roy} {and}
  \bibinfo{person}{Sharath~Chandra Guntuku}.} \bibinfo{year}{2016}\natexlab{}.
\newblock \showarticletitle{Latent factor representations for cold-start video
  recommendation}. In \bibinfo{booktitle}{\emph{RecSys}}. ACM,
  \bibinfo{pages}{99--106}.
\newblock


\bibitem[\protect\citeauthoryear{Saveski and Mantrach}{Saveski and
  Mantrach}{2014}]%
        {saveski2014item}
\bibfield{author}{\bibinfo{person}{Martin Saveski} {and} \bibinfo{person}{Amin
  Mantrach}.} \bibinfo{year}{2014}\natexlab{}.
\newblock \showarticletitle{Item cold-start recommendations: learning local
  collective embeddings}. In \bibinfo{booktitle}{\emph{RecSys}}. ACM,
  \bibinfo{pages}{89--96}.
\newblock


\bibitem[\protect\citeauthoryear{Schein, Popescul, Ungar, and Pennock}{Schein
  et~al\mbox{.}}{2002}]%
        {schein2002methods}
\bibfield{author}{\bibinfo{person}{Andrew~I Schein},
  \bibinfo{person}{Alexandrin Popescul}, \bibinfo{person}{Lyle~H Ungar}, {and}
  \bibinfo{person}{David~M Pennock}.} \bibinfo{year}{2002}\natexlab{}.
\newblock \showarticletitle{Methods and metrics for cold-start
  recommendations}. In \bibinfo{booktitle}{\emph{SIGIR}}. ACM,
  \bibinfo{pages}{253--260}.
\newblock


\bibitem[\protect\citeauthoryear{Seroussi, Bohnert, and Zukerman}{Seroussi
  et~al\mbox{.}}{2011}]%
        {seroussi2011personalised}
\bibfield{author}{\bibinfo{person}{Yanir Seroussi}, \bibinfo{person}{Fabian
  Bohnert}, {and} \bibinfo{person}{Ingrid Zukerman}.}
  \bibinfo{year}{2011}\natexlab{}.
\newblock \showarticletitle{Personalised rating prediction for new users using
  latent factor models}. In \bibinfo{booktitle}{\emph{HT}}. ACM,
  \bibinfo{pages}{47--56}.
\newblock


\bibitem[\protect\citeauthoryear{Shah, Yang, Alle, Ratnaparkhi, Shahshahani,
  and Chandra}{Shah et~al\mbox{.}}{2017}]%
        {shah2017practical}
\bibfield{author}{\bibinfo{person}{Parikshit Shah}, \bibinfo{person}{Ming
  Yang}, \bibinfo{person}{Sachidanand Alle}, \bibinfo{person}{Adwait
  Ratnaparkhi}, \bibinfo{person}{Ben Shahshahani}, {and} \bibinfo{person}{Rohit
  Chandra}.} \bibinfo{year}{2017}\natexlab{}.
\newblock \showarticletitle{A practical exploration system for search
  advertising}. In \bibinfo{booktitle}{\emph{KDD}}. ACM,
  \bibinfo{pages}{1625--1631}.
\newblock


\bibitem[\protect\citeauthoryear{Song, Shi, Xiao, Duan, Xu, Zhang, and
  Tang}{Song et~al\mbox{.}}{2019}]%
        {song2019autoint}
\bibfield{author}{\bibinfo{person}{Weiping Song}, \bibinfo{person}{Chence Shi},
  \bibinfo{person}{Zhiping Xiao}, \bibinfo{person}{Zhijian Duan},
  \bibinfo{person}{Yewen Xu}, \bibinfo{person}{Ming Zhang}, {and}
  \bibinfo{person}{Jian Tang}.} \bibinfo{year}{2019}\natexlab{}.
\newblock \showarticletitle{Autoint: Automatic feature interaction learning via
  self-attentive neural networks}. In \bibinfo{booktitle}{\emph{CIKM}}. ACM,
  \bibinfo{pages}{1161--1170}.
\newblock


\bibitem[\protect\citeauthoryear{Tang, Jiang, Li, Zeng, and Li}{Tang
  et~al\mbox{.}}{2015}]%
        {tang2015personalized}
\bibfield{author}{\bibinfo{person}{Liang Tang}, \bibinfo{person}{Yexi Jiang},
  \bibinfo{person}{Lei Li}, \bibinfo{person}{Chunqiu Zeng}, {and}
  \bibinfo{person}{Tao Li}.} \bibinfo{year}{2015}\natexlab{}.
\newblock \showarticletitle{Personalized recommendation via parameter-free
  contextual bandits}. In \bibinfo{booktitle}{\emph{SIGIR}}. ACM,
  \bibinfo{pages}{323--332}.
\newblock


\bibitem[\protect\citeauthoryear{Vartak, Thiagarajan, Miranda, Bratman, and
  Larochelle}{Vartak et~al\mbox{.}}{2017}]%
        {vartak2017meta}
\bibfield{author}{\bibinfo{person}{Manasi Vartak}, \bibinfo{person}{Arvind
  Thiagarajan}, \bibinfo{person}{Conrado Miranda}, \bibinfo{person}{Jeshua
  Bratman}, {and} \bibinfo{person}{Hugo Larochelle}.}
  \bibinfo{year}{2017}\natexlab{}.
\newblock \showarticletitle{A Meta-Learning Perspective on Cold-Start
  Recommendations for Items}. In \bibinfo{booktitle}{\emph{NIPS}}.
\newblock


\bibitem[\protect\citeauthoryear{Veli{\v{c}}kovi{\'c}, Cucurull, Casanova,
  Romero, Li{\`o}, and Bengio}{Veli{\v{c}}kovi{\'c} et~al\mbox{.}}{2018}]%
        {velivckovic2018graph}
\bibfield{author}{\bibinfo{person}{Petar Veli{\v{c}}kovi{\'c}},
  \bibinfo{person}{Guillem Cucurull}, \bibinfo{person}{Arantxa Casanova},
  \bibinfo{person}{Adriana Romero}, \bibinfo{person}{Pietro Li{\`o}}, {and}
  \bibinfo{person}{Yoshua Bengio}.} \bibinfo{year}{2018}\natexlab{}.
\newblock \showarticletitle{Graph Attention Networks}. In
  \bibinfo{booktitle}{\emph{ICLR}}.
\newblock


\bibitem[\protect\citeauthoryear{Volkovs, Yu, and Poutanen}{Volkovs
  et~al\mbox{.}}{2017}]%
        {volkovs2017dropoutnet}
\bibfield{author}{\bibinfo{person}{Maksims Volkovs}, \bibinfo{person}{Guang~Wei
  Yu}, {and} \bibinfo{person}{Tomi Poutanen}.} \bibinfo{year}{2017}\natexlab{}.
\newblock \showarticletitle{DropoutNet: Addressing Cold Start in Recommender
  Systems.}. In \bibinfo{booktitle}{\emph{NIPS}}. \bibinfo{pages}{4957--4966}.
\newblock


\bibitem[\protect\citeauthoryear{Wang, Fu, Fu, and Wang}{Wang
  et~al\mbox{.}}{2017}]%
        {wang2017deep}
\bibfield{author}{\bibinfo{person}{Ruoxi Wang}, \bibinfo{person}{Bin Fu},
  \bibinfo{person}{Gang Fu}, {and} \bibinfo{person}{Mingliang Wang}.}
  \bibinfo{year}{2017}\natexlab{}.
\newblock \showarticletitle{Deep \& cross network for ad click predictions}. In
  \bibinfo{booktitle}{\emph{ADKDD}}. ACM, \bibinfo{pages}{12}.
\newblock


\bibitem[\protect\citeauthoryear{Xiong, Wang, Ding, Shen, and Liu}{Xiong
  et~al\mbox{.}}{2012}]%
        {xiong2012relational}
\bibfield{author}{\bibinfo{person}{Chenyan Xiong}, \bibinfo{person}{Taifeng
  Wang}, \bibinfo{person}{Wenkui Ding}, \bibinfo{person}{Yidong Shen}, {and}
  \bibinfo{person}{Tie-Yan Liu}.} \bibinfo{year}{2012}\natexlab{}.
\newblock \showarticletitle{Relational click prediction for sponsored search}.
  In \bibinfo{booktitle}{\emph{WSDM}}. ACM, \bibinfo{pages}{493--502}.
\newblock


\bibitem[\protect\citeauthoryear{Xu, Liu, Shu, and Yu}{Xu
  et~al\mbox{.}}{2018}]%
        {xu2018lifelong}
\bibfield{author}{\bibinfo{person}{Hu Xu}, \bibinfo{person}{Bing Liu},
  \bibinfo{person}{Lei Shu}, {and} \bibinfo{person}{Philip~S Yu}.}
  \bibinfo{year}{2018}\natexlab{}.
\newblock \showarticletitle{Lifelong domain word embedding via meta-learning}.
  In \bibinfo{booktitle}{\emph{IJCAI}}. \bibinfo{pages}{4510--4516}.
\newblock


\bibitem[\protect\citeauthoryear{Yin, Mei, Cao, Sun, and Davison}{Yin
  et~al\mbox{.}}{2014}]%
        {yin2014exploiting}
\bibfield{author}{\bibinfo{person}{Dawei Yin}, \bibinfo{person}{Shike Mei},
  \bibinfo{person}{Bin Cao}, \bibinfo{person}{Jian-Tao Sun}, {and}
  \bibinfo{person}{Brian~D Davison}.} \bibinfo{year}{2014}\natexlab{}.
\newblock \showarticletitle{Exploiting contextual factors for click modeling in
  sponsored search}. In \bibinfo{booktitle}{\emph{WSDM}}. ACM,
  \bibinfo{pages}{113--122}.
\newblock


\bibitem[\protect\citeauthoryear{Zhang, Tang, Zhang, and Xue}{Zhang
  et~al\mbox{.}}{2014}]%
        {zhang2014addressing}
\bibfield{author}{\bibinfo{person}{Mi Zhang}, \bibinfo{person}{Jie Tang},
  \bibinfo{person}{Xuchen Zhang}, {and} \bibinfo{person}{Xiangyang Xue}.}
  \bibinfo{year}{2014}\natexlab{}.
\newblock \showarticletitle{Addressing cold start in recommender systems: A
  semi-supervised co-training algorithm}. In \bibinfo{booktitle}{\emph{SIGIR}}.
  ACM, \bibinfo{pages}{73--82}.
\newblock


\bibitem[\protect\citeauthoryear{Zhang, Du, and Wang}{Zhang
  et~al\mbox{.}}{2016}]%
        {zhang2016deep}
\bibfield{author}{\bibinfo{person}{Weinan Zhang}, \bibinfo{person}{Tianming
  Du}, {and} \bibinfo{person}{Jun Wang}.} \bibinfo{year}{2016}\natexlab{}.
\newblock \showarticletitle{Deep learning over multi-field categorical data}.
  In \bibinfo{booktitle}{\emph{ECIR}}. Springer, \bibinfo{pages}{45--57}.
\newblock


\bibitem[\protect\citeauthoryear{Zhou, Mou, Fan, Pi, Bian, Zhou, Zhu, and
  Gai}{Zhou et~al\mbox{.}}{2019}]%
        {zhou2019deep}
\bibfield{author}{\bibinfo{person}{Guorui Zhou}, \bibinfo{person}{Na Mou},
  \bibinfo{person}{Ying Fan}, \bibinfo{person}{Qi Pi}, \bibinfo{person}{Weijie
  Bian}, \bibinfo{person}{Chang Zhou}, \bibinfo{person}{Xiaoqiang Zhu}, {and}
  \bibinfo{person}{Kun Gai}.} \bibinfo{year}{2019}\natexlab{}.
\newblock \showarticletitle{Deep interest evolution network for click-through
  rate prediction}. In \bibinfo{booktitle}{\emph{AAAI}},
  Vol.~\bibinfo{volume}{33}. \bibinfo{pages}{5941--5948}.
\newblock


\bibitem[\protect\citeauthoryear{Zhou, Zhu, Song, Fan, Zhu, Ma, Yan, Jin, Li,
  and Gai}{Zhou et~al\mbox{.}}{2018}]%
        {zhou2018deep}
\bibfield{author}{\bibinfo{person}{Guorui Zhou}, \bibinfo{person}{Xiaoqiang
  Zhu}, \bibinfo{person}{Chenru Song}, \bibinfo{person}{Ying Fan},
  \bibinfo{person}{Han Zhu}, \bibinfo{person}{Xiao Ma},
  \bibinfo{person}{Yanghui Yan}, \bibinfo{person}{Junqi Jin},
  \bibinfo{person}{Han Li}, {and} \bibinfo{person}{Kun Gai}.}
  \bibinfo{year}{2018}\natexlab{}.
\newblock \showarticletitle{Deep interest network for click-through rate
  prediction}. In \bibinfo{booktitle}{\emph{KDD}}. ACM,
  \bibinfo{pages}{1059--1068}.
\newblock


\bibitem[\protect\citeauthoryear{Zhou, Yang, and Zha}{Zhou
  et~al\mbox{.}}{2011}]%
        {zhou2011functional}
\bibfield{author}{\bibinfo{person}{Ke Zhou}, \bibinfo{person}{Shuang-Hong
  Yang}, {and} \bibinfo{person}{Hongyuan Zha}.}
  \bibinfo{year}{2011}\natexlab{}.
\newblock \showarticletitle{Functional matrix factorizations for cold-start
  recommendation}. In \bibinfo{booktitle}{\emph{SIGIR}}. ACM,
  \bibinfo{pages}{315--324}.
\newblock


\end{thebibliography}

\end{document}